\documentclass[5p,times]{elsarticle}

\usepackage{amssymb}
\usepackage{amsmath}

\journal{Future Generation Computer Systems}

\usepackage{comment}
\usepackage{url}
\usepackage{float}
\usepackage{hyperref}
\usepackage{array}
\usepackage{booktabs}
\usepackage{caption}
\usepackage{longtable}
\usepackage{tabularx}

\usepackage{multirow}
\usepackage{subcaption}
\usepackage{multicol}
\usepackage{tablefootnote}
\usepackage{enumitem}

\newcommand{\Description}[1]{}

\newcommand{\systemName}{\textit{EcoMate}}

\begin{document}

\begin{frontmatter}



\title{Generating HomeAssistant Automations Using an LLM-based Chatbot}


\author[polimi]{Mathyas Giudici\corref{cor}} 
\author[polimi]{Alessandro Sironi}
\author[polimi]{Ismaele Villa}
\author[polimi]{Samuele Scherini}
\author[polimi,unimib]{Franca Garzotto}

\cortext[cor]{Corresponding Author [name.surname@polimi.it]}

\affiliation[polimi]{organization={Department of Electronics, Information, and Bioengineering (DEIB), Politecnico di Milano},
            addressline={Piazza Leonardo Da Vinci, 32}, 
            city={Milano},
            postcode={20133}, 
            state={Italy},
            country={}}

\affiliation[unimib]{organization={Department of Psychology,  Università degli Studi di Milano-Bicocca},
            addressline={Piazza dell'Ateneo Nuovo, 1}, 
            city={Milano},
            postcode={20126}, 
            state={Italy},
            country={}}

\begin{abstract}
To combat climate change, individuals are encouraged to adopt sustainable habits, in particular, with their household, optimizing their electrical consumption. Conversational agents, such as Smart Home Assistants, hold promise as effective tools for promoting sustainable practices within households.
Our research investigated the application of Large Language Models (LLM) in enhancing smart home automation and promoting sustainable household practices, specifically using the HomeAssistant framework. In particular, it highlights the potential of GPT models in generating accurate automation routines. While the LLMs showed proficiency in understanding complex commands and creating valid JSON outputs, challenges such as syntax errors and message malformations were noted, indicating areas for further improvement. Still, despite minimal quantitative differences between "green" and "no green" prompts, qualitative feedback highlighted a positive shift towards sustainability in the routines generated with environmentally focused prompts.
Then, an empirical evaluation (N=56) demonstrated that the system was well-received and found engaging by users compared to its traditional rule-based counterpart.
Our findings highlight the role of LLMs in advancing smart home technologies and suggest further research to refine these models for broader, real-world applications to support sustainable living.
\end{abstract}





\end{frontmatter}



\section{Introduction}
\label{sec:intro}

Individual energy consumption habits are crucial in achieving sustainability and mitigating climate change \cite{abdullah2022individual,allen2018special}.
The residential electricity consumption accounts for a significant portion of the world's energy consumption~\cite{iea2022world}. In addition, such consumption will increase since electrification is emerging as a crucial strategy for sustainable development \cite{stewart2018electric}.

Interactive and innovative technologies and Human-Computer Interaction research can play a pivotal role in providing new digital solutions and addressing the challenge of raising awareness and guiding individuals toward more sustainable behaviors~\cite{vinuesa2020role,disalvo2010mapping}.
It has been demonstrated that digital systems that offer sustainability and environmental-related feedback, such as real-time energy consumption visualization, promote mindful use of electricity.
Home automation (grounded on IoT networks) enables the monitoring of house appliances and can help promote strategies to optimize their usage~\cite{giudici2022candy, hussain2019survey}. Home assistants, such as Alexa or Google Home (i.e. conversational agents), represent one of the most popular interfaces for interacting with domestic IoT networks~\cite{sciuto2018hey}, using the so-called Trigger-Action Programming~\cite{corno2020heytap} (i.e., the definition of simple programs to trigger home appliances in specified time or sensed conditions).
However, non-expert users tend to rely on default settings, while more advanced users attempting customization found difficulties in translating their intentions into automation \cite{gallo2024conversational}. Still, the need for continual adjustments to accommodate changing preferences further complicates the user experience.

The emergence of Large Language Models (LLMs) in AI has transformed a number of conversational technology applications, such as code development, story and creative generation, and text summarization~\cite{raiaan2024review}. Although LLMs have found use in a variety of fields, including robotics~\cite{wu2023tidybot} and video games~\cite{park2023generative}, there is still much to learn about how to integrate LLMs into smart home automation~\cite{king2023sasha,gallo2024conversational}, especially in the area of environmental sustainability.
For example, \citet{king2023sasha} investigated the usage of LLM in a home automation environment by specifying in the prompt of the LLM the specific JSON format for the set of devices they had in their experimental setting.

In our research, we introduced and contributed to the usage of LLM-based conversational agents in sustainable smart homes, addressing the \emph{research gap between the implicit goals of users in smart homes, the appliances' activities required to achieve them, and the implementation into HomeAssistant\footnote{\url{https://www.home-assistant.io/}}, a popular open-source home automation framework}.
To investigate such a gap, we postulated the following research questions:

\begin{enumerate}[label=\textbf{RQ\arabic*}]
    \item \emph{Are LLMs able to control smart home environments using the HomeAssistant framework?} What are the challenges, potentially generated inaccuracies, or "hallucinations" to be addressed?

    \item \emph{Are LLMs able to control smart home environments more sustainably?} Is there a difference between "green" vs. "no green" generated actions? Does a more "green" prompt potentially generate inaccuracies or "hallucinations"?

    \item \emph{How possible end users perceive such LLM-based home assistants compared to traditional rule-based chatbots?} What are the user engagement, likability, and usability?
\end{enumerate}

\begin{figure}[ht]
    \includegraphics[width=\linewidth]{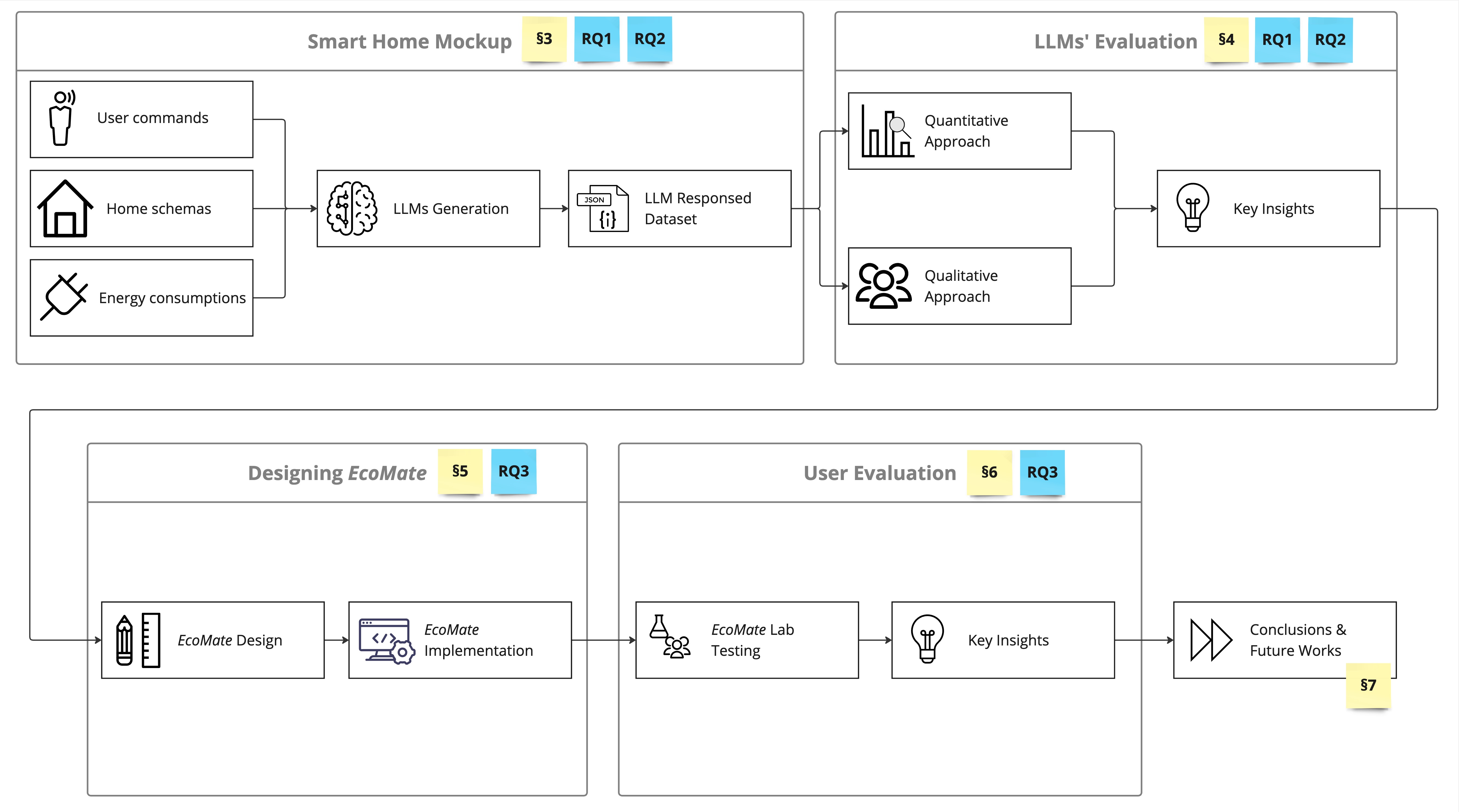}
    \caption{Paper Overview, with mapping of RQs to empirical studies and sections}
    \Description{Paper Overview, with mapping of RQs to empirical studies and sections}
    \label{figure:paper_overview}
\end{figure}

Figure \ref{figure:paper_overview} describes the overall research approach. First, we investigated the capabilities of different LLMs in generating HomeAssistant home automation for controlling smart home environments.
We developed a prototype that combines in the prompt: natural language user commands along with the "home template" and the energy consumption of home appliances (all inspired by literature datasets).
The results showed that GPT models are the best LLMs in the generation of HomeAssistant routines. However, there are some challenges in providing accurate JSON formatting and correctly interpreting user commands. A comparison of a prompt that promotes more environmentally sustainable behavior did not show any particular impactful differences in routine generation compared to a "no green" prompt. 

Then, we implemented \textit{EcoMate}, a GPT3.5-powered chatbot, to guide users toward sustainable energy usage, enable routine generation, and ensure seamless integration with HomeAssistant.
To evaluate the effectiveness of \textit{EcoMate}, we conducted an empirical study involving 56 participants and assessed the chatbot engagement and usability capabilities compared to its rule-based counterpart. Users perceived the system as usable, while interaction satisfaction and perceived dialogue were statistically greater for the LLM-based version, which remarks its potential for encouraging sustainable practices in home environments.
Our findings offer valuable insights into LLMs' effectiveness and user acceptance, paving the way for further exploration of LLM-based HomeAssistant controllers that can also be used to promote environmentally sustainable behaviors within home settings.

The paper is organized as follows. Section \ref{sec:soa} presents the research landscape. 
Section \ref{sec:home-mockup} presents the first study setup, describing the home environments selected, user actions, energy data, and LLMs under evaluation. Section \ref{sec:llm-comparation} discusses the results of the comparison of the different LLMs. Section \ref{sec:ecomate-design} describes the design of \systemName{}, while Section \ref{sec:ecomate-user-eval} its user evaluation and result discussion. Finally, Section \ref{sec:conclusion} summarises and concludes the implications of the proposed research work, also offering possible future works to implement in the field.

\section{State of the Art}
\label{sec:soa}

\subsection{Home Automation Environments}
Smart environments are physical spaces enhanced with sensing, actuation, communication, and computation capabilities to adapt to users' preferences, requirements, and specific needs \cite{cicirelli2016meta}. They encompass various applications and settings, from smart homes to smart cities and factories \cite{el2021smart}.
The development of smart environments faces challenges such as precise activity recognition, effective localization systems, and the need for a smooth transition from traditional to smart environments \cite{degeler2014architecture,evangelatos2013syndesi}. Emerging wearable devices and wireless communication technologies are expected to further optimize the resource efficiency, comfort, and safety of such environments \cite{el2021smart}.

In this wide landscape, \emph{home automation} environments refer to systems that enable the management of household activities through computerized control, offering benefits such as comfort, security, and energy efficiency~\cite{yuneela2022review}. Advancements in the Internet of Things (IoT) research enabled remote control of devices and simplified everyday tasks, making home automation -- in general -- more accessible and user-friendly, using web/mobile applications or home automation systems~\cite{omran2021survey}. In order to control home appliances, Trigger-Action-Programming (TAP) is a simple programming model available to users to create rules automating the behavior of smart homes, devices, and online services \cite{ur2014practical}. Householders can create TAP rules (also called recipes, applets, or routines) using a simple conditional structure -- summarised in \texttt{if 'condition' then 'action'} -- using applications designed to be user-friendly and accessible without a programming background \cite{10.1145/2858036.2858556}. For instance, commercial services, such as IFTTT\footnote{\url{https://ifttt.com/}} (i.e., If This Then That) and Zapier\footnote{\url{https://zapier.com/}}, enable end-users to simply use TAP with the largest commercially available home appliances. Still, there are open-source solutions~\cite{khusnutdinov2018open} that allow full home automation control. One example is  HomeAssistant. As presented by \citet{akhmetzhanov2022integration}, it was used to create a video surveillance system using different commercial cameras.

However, according to \cite{chen2022fix, huang2015supporting}, the oversimplification of existing TAP systems limits the expressivity of the programs that can be created, leading to inconsistencies in interpreting the behavior of TAP and errors in creating programs with a desired behavior. In addition, another limitation is the lack of standard open protocols: each vendor implements its own product in a closed environment and requires the user to use several applets to manage a fully integrated environment \cite{corno2019recrules}.
Still, empirical evidence from \citet{10.1145/2858036.2858556} reported how users tend to create their TAP recipes rather than using existing ones shared by other users or appliance manufacturers. Finally, \citet{heo2017rt} discussed how existing IoT frameworks read sensor data periodically, independent of real-time constraints. For example, the IFTTT framework polls sensor data every 15 minutes, and real-time APIs cause sensors to send unnecessary messages that do not affect any TAP recipe and waste battery power. In their work, the authors introduced the RT-IFTTT language, which allows users to specify real-time sensor constraints. A central manager analyses the relationships between sensors and the TAP, calculating polling intervals for each trigger condition and disabling the sensor when not needed, saving battery and energy.

While TAP has demonstrated great practicality in customizing smart home devices, allowing end users to express a wide range of desired behaviors, and machine learning algorithms have improved the adaptability and functionality of such programs \cite{ur2014practical}, the landscape is still missing in the sustainable living field.
However, it is worth reporting that in a broader panorama, as in the Sustainable Human-Computer Interaction~\cite{disalvo2010mapping} literature, we can find the evaluation of different visual systems~\cite{froehlich2010design} to reduce the energy consumption of households~\cite{schwartz2013uncovering, pierce2012beyond,costanza2012understanding} or intelligent technologies that shift people’s thermal comfort consumption~\cite{clear2014catch}. Still in this field, \citet{alan2016tariff} presented an interactive IoT system designed to help manage energy costs, offering flexible autonomy and detailed information about its operation.
Finally, \citet{bang2006powerhhouse} proposed a gamified activity in the context of domestic sustainability, while \citet{beheshtian2020greenlife} developed a social robot for sustainable living in a block of flats.
These examples from the Sustainable HCI field can inform the design of new TAP solutions to help people have more sustainable home environments (e.g., designing applets to reduce energy consumption).

\subsection{Large Language Models}
A \textit{Large Language Model (LLM)} is a subset of Artificial Intelligence algorithms that uses deep learning techniques (e.g., transformer model) and massively large datasets to understand, summarize, generate and predict new content~\cite{vaswani2017attention, raiaan2024review}.

Nowadays, LLMs are mainly employed for text summarization~\cite{hoang2019efficient} and Next Sequence Prediction~\cite{liu2019roberta} (i.e., predict the subsequent element or event in a sequence based on patterns and information present in the history of the sequence).
%
Large Language Models are grounded upon extensive scientific research in Natural Language Processing, aiming to create computers that are able to interact with the user using Natural Language~\cite{hadi2023survey} (i.e., interpret and generate human language).
A digital tool that interacts with users using natural language is defined in scientific literature as a \emph{Conversational Agent (CA)}~\cite{masche2017review,rheu2021systematic}. Historically, they have been used to engage users in text-based information-seeking and task-oriented dialogues for many applications~\cite{lester2004conversational}. For example, they are integrated into physical devices (such as Alexa and Google Home \cite{sciuto2018hey,bentley2018understanding}) and available in many contexts of everyday life, used in phones (like Siri, the Apple virtual assistant \cite{jaber2020conversational}), cars, and online consumer assistance~\cite{bavaresco2020conversational}.

CAs are also considered promising in the environmental sustainability domain \cite{giudici2022candy, hussain2019survey}, considering that they are already prevalent in numerous households as home assistants~\cite{sciuto2018hey,10.1145/3469595.3469602}, and integrated with IoT appliances~\cite{10.1145/2858036.2858518,fischer2017data} and TAP\cite{corno2020heytap} to promote sustainability in the domestic settings. Traditional rule-based chatbots have previously been used to deliver energy feedback~\cite{gnewuch2018designing,10.1145/3582515.3609526} -- also using IFTTT~\cite{biswas2023sustainable}, suggesting sustainable mobility~\cite{diederich2019promoting}, or reducing food waste~\cite{cacanindin_greening_2020}, and in the school education context~\cite{rukhiran2022adoption}. Still, \citet{ramasubbu2018intrusive} presented a chatbot to optimize the schedule of switching off smart plugs in an office. 
Instead, \citet{gunawardane2019zero} proposed an example using a data-driven chatbot to suggest recipes with leftover foods. Finally, \citet{giudici2023assessing} presented an evaluation of LLMs to create a hybrid conversational agent able to trigger devices in a home automation environment and address open-domain questions by users.

Although rule-based CAs have been a milestone in smart home conversational interaction, they have significant limitations, such as a lack of adaptability and scalability of conversations~\cite{thorat2020review}. On the contrary, LLMs unlock a more natural interaction between user and space compared to the interaction that task-specific systems can provide~\cite{king2023sasha}. 
In addition, they can overcome the limitations of traditional conversational agents integrated with TAP, which are a limited language pattern to follow, triggering and interfacing with single devices, leading to a low integrated ecosystem perception \cite{mi2017empirical}.
This is possible thanks to a major exposure of LLM to daily situations and requests inserted into their diverse training data,  or thanks to fine-tuning techniques (i.e., using previously available datasets~\cite{noura2020vish}).
For instance, \citet{fast2016augur} presented that an LLM trained only in written works of fiction is able to recognize everyday activities based on semantic relationships between objects and the activities they are frequently used in. 
\citet{nascimento2023artificial} explored the usage of generative AI versus engineering-crafted coding to create a control structure for an IoT application. Results pointed out how there were cases in which humans outperformed AI algorithms and others where they didn’t, with the experiment validity highly impacted by the background experience and skills of the engineers enrolled. Still, \citet{li2023chatiot} presented ChatIoT, a zero-code rule generator, to improve the quality of TAP generation while reducing the tokens required for the prompt to the LLM. In addition, a set of confirming rules in the interaction pipeline allows refining the TAP recipe to be more accurate and safely executed.
Finally, relevant to our research is the approach proposed by \citet{king2023sasha}, who presented \emph{Sasha}, an LLM-based application able to create goal-oriented home automation in domotic environments. They evaluated the ability of such LLM to create parsable JSON files to be applied in a real setting using a data source of common smart home action plans. 

To the best of our knowledge, an aspect that remains unexplored is the utilization of LLMs to guide users in fostering sustainability within their homes. Specifically, we focus on evaluating LLMs' models to assist households in creating routines that promote eco-sustainability and energy conservation directly sent and integrated into a running instance of the HomeAssistant environment.

\section{Comparative Study Setup}
\label{sec:home-mockup}
We crafted a prototype using LLMs to generate action plans in smart homes responding to user commands to run our comparative study and address the first two research questions (i.e., RQ1 and RQ2 as defined in Section \ref{sec:intro}).
Our system sends prompt (using the zero-shot prompt engineering technique) commands to an LLM including a "home template" in JSON that lists the rooms, appliances, and sensors in a smart home, the energy consumption of the appliances in the home, and the user's natural language command.
Finally, to better support our empirical work, many of the experimental choices were based on data and configurations found in the literature (and detailed in the following subsections), but we aim to make a strong and novel contribution by addressing the generation of both "green" and "no green" home automations, directly formatted for the HomeAssistant framework API.

\subsection{Prompt}
\label{sec:home-mockup|subsec:prompt}

The LLMs' model size and the inaccessibility of resources and constraints on which they are tuned made emerging the \emph{prompt engineering}~\cite{zhao2021calibrate, zhang2022promote}, a set of techniques which involve methods utilized to control the LLMs' inputs and influence their output. This methodology leverages describing (zero-shot) or presenting examples (N-shot) of a given task to a pre-trained LLM model without the need for fine-tuning~\cite{brown2020language}. Notably, for the aim of our research, avoiding the fine-tuning of models also represents an environmentally sustainable choice, as the resource-intensive GPU training often results in substantial carbon dioxide emissions, as highlighted in the work of \citet{strubell2019energy}.

The zero-shot prompt includes different instructions (hereinafter \textit{Part}) provided to the LLM for each question the users submit. In addition, a generic prompt ("no green") and one more oriented to environmental sustainability ("green") were created to address the two research questions. In particular, the "green" prompt clarifies that the routines aim to address sustainable energy practices, energy consumption optimization, and cultivating environmentally friendly habits. \ref{app:comparison-promt} provides the complete text of the two prompts.

Both the prompts include a \emph{General Part} to define the chatbot behavior and provide the context in which the model operates, a \emph{Routine Part} specifying conditions in formatting HomeAssistant routines, and an \emph{Explanation Part} to define insights into the response methodology to follow. The prompt is prepended with the smart home configuration and appliances' energy consumption.
Following such an initial prompt, the LLM usually generates a response with a JSON structure parsable by HomeAssistant API. Since the model responses are in a MarkDown format, the generated JSON object is usually encapsulated within three backtick symbols (i.e., \texttt{\textasciigrave\textasciigrave\textasciigrave}). Thus, the system logic locates it and automatically integrates the routine into the home automation system. 

\subsection{User Commands}
As a user commands dataset, we used the 40 natural language utterances of typical smart home user requests created by \citet{king2023sasha} (that was based on the work by \citet{yu2021analysis}).
The dataset comprises seven categories targeting groups of user requests based on the types of actions and devices they refer to. Each command is also characterized by an example of an IFTTT routine and the type of goal (immediate or persistent).
For example, the "Ambient Luminance" category conveys the user's intention to adjust the lighting, and a user command within that category is "make it less bright" which is associated with the "Change brightness of your light" IFTTT and has an immediate goal.

\subsection{Smart Homes}
Smart homes' configurations were adapted starting from the CASAS dataset~\cite{cook2012casas,cook2010learning} (available online\footnote{\url{https://casas.wsu.edu/datasets/}}). 
The dataset provides different configurations of smart flats with different numbers of tenants and each apartment is equipped with various motion sensors distributed throughout the space. It is one of the most comprehensive and widely used datasets containing house templates in the literature~\cite{dahmen2017smart}.

\begin{figure}
    \centering
    \begin{subfigure}[t]{.75\linewidth}
        \centering
        \includegraphics[width=\linewidth]{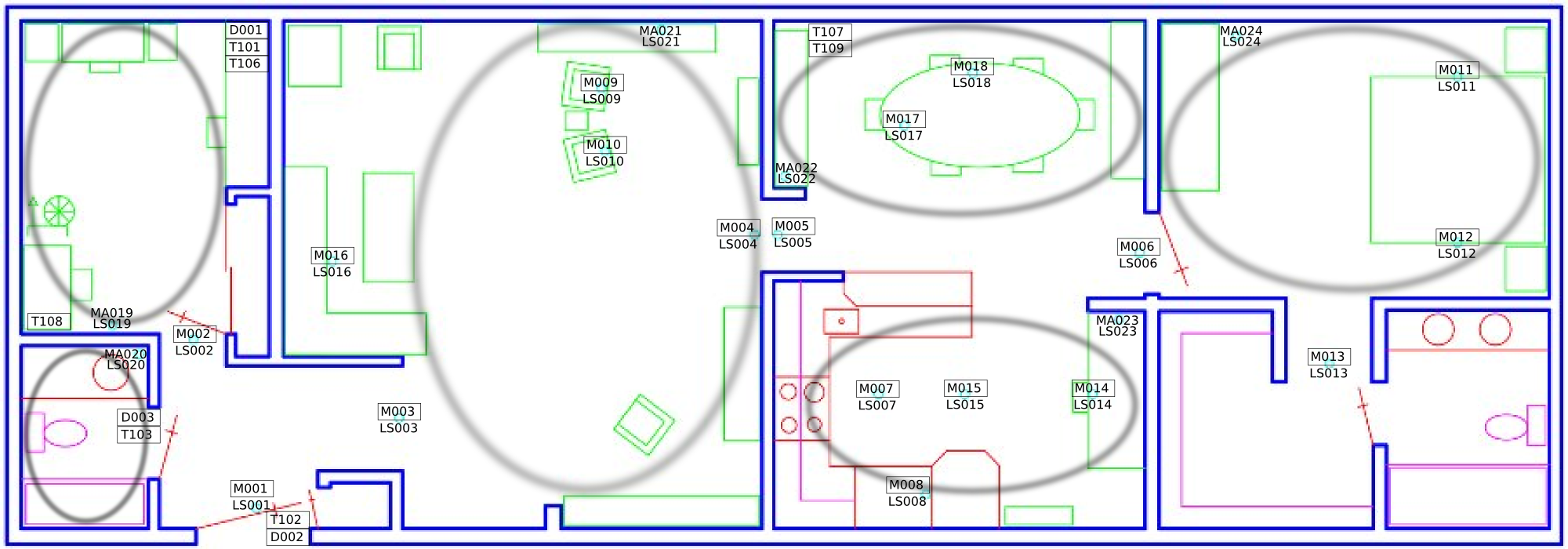}
        \caption{Original H107 House Template}
        \label{figure:home-templates|sub-ori}
    \end{subfigure}
    \begin{subfigure}[t]{.75\linewidth}
        \centering
        \includegraphics[width=\linewidth]{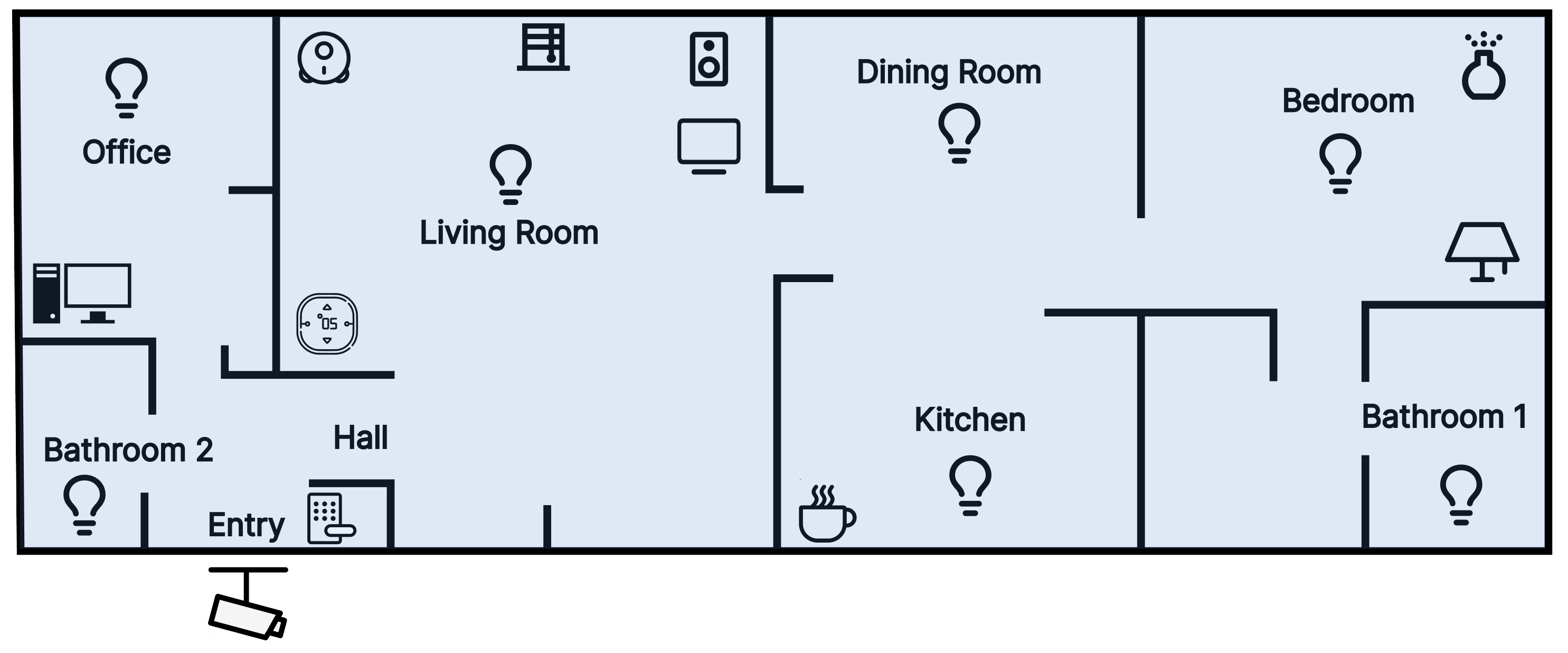}
        \caption{Revisited H107 House Template (including appliances)}
        \label{figure:home-templates|sub-rev}
    \end{subfigure}
    \caption{Original and Revisited H107 House Templates}
    \label{figure:home-templates}
    \Description{Pictures of original and revised H107 house template from CASAs dataset}
\end{figure}

For our research setting, we selected house H107 based on two-resident apartment data and defined with a specific template (see Figure \ref{figure:home-templates|sub-ori}). In addition, the provided home configurations (in the CASAS dataset) only refer to movement sensors and lighting. To have an experimental setting close to a real-case scenario (as also done in the previous work by \citet{king2023sasha}), we added common appliances in houses, as depicted in Figure \ref{figure:home-templates|sub-rev}. The set of appliances consists of lightbulbs, a coffee machine, speakers, a television, smart blinds, an air purifier, a smart lock, a security camera, and a vacuum cleaner.

\subsection{Energy Consumption}
Home appliance energy consumption was based on the data collected by the work of \citet{panagiotidou2023supporting}. They explored the usage of a sensor kit in 14 UK households to monitor and contextually annotate the participants' electricity consumption over a period of 5-15 days.

For our context, we concatenated the different annotation datasets of the participants, extracted the appliance types, and computed the average consumption within the sampling period. 
Since some pre-cleaning process of the folder containing the dataset is needed and to enable other people in the scientific community to replicate our study, the code to manipulate the annotation is provided\footnote{\url{https://gitlab.com/i3lab/ecomate/energyconsumption-preparation}}.

\subsection{Methodology}

\subsubsection{Study Design}
The study follows a between-factors design: 6 LLMs x 2 temperatures (t) x 2 prompts (i.e., "green" and "no green").
We selected the most popular and available LLM models at the time of the study (i.e., early 2024), and they are GPT3.5-Turbo, GPT4, LLAMA2-7b, LLAMA2-70b, MISTRAL (8x7B), and codeLLAMA (7B). GPT3.5-Turbo was accessed using Azure OpenAI Service, while for GPT4 OpenAI API service was used. The other models were accessed by running their instances on Azure AI Studio Services Virtual machines (Standard\_NC12s) and exposing live endpoints. In all the cases, we access the APIs from a 1Gbps fiber internet connection.
According to previous work by \citet{king2023sasha}, we picked two different temperatures for the model, 0 and 0.7 respectively. Finally, as described in Section \ref{sec:home-mockup|subsec:prompt}, two different prompts were used to guide the generation.

\subsubsection{Metrics}
\label{sec:home-mockup|subsec:metod-metrics}
Data gathering was performed using custom code, as described above. Then, metrics were computed to evaluate the performance of the LLMs under the different design conditions when generating action plans.
We relied on metrics described in \citet{king2023sasha}, which are defined by the following terms:
\begin{itemize}
    \item \textbf{False positives (FP)} is defined as a routine that targets one or more devices, but the home does not contain any relevant appliance.
    \item \textbf{False negatives (FN)} represents when the LLM does not produce a routine, but the home contains important appliances to build it.
    \item \textbf{Accuracy (Acc)} is defined with the formula $Acc = 1 - (FP + FN)$.
    \item \textbf{Relevance score (Rel)} is a score (ranging between -1 and 1) that, for each command, highlights how many relevant and irrelevant devices are used to create the routine.
    \item \textbf{Latency} (in seconds) is determined by timing the call of the LLM model API.
\end{itemize}

However, other metrics were modified and computed as described:
\begin{itemize}
    \item \textbf{JSON validity.} We automatically extracted the JSON code of the home automation (see Section \ref{sec:home-mockup|subsec:prompt}) and sent it to the HomeAssistant framework via API. In the case of a positive response, the validity was flagged as "true", whereas was marked as "false" for the negative ones. In the latter case, the error raised by the framework was also noted.
    \item \textbf{JSON failure type.} We classified the failed routines (i.e., with "false" \emph{JSON validity}) according to the groups defined in \citet{king2023sasha}\footnote{See Section 4.4.2 of the cited work for more details}. The clusters previously identified were seven; we added an eighth group ("other"), including all the failures that do not map with such classification.  Additionally, we analyzed such "other" category to extrapolate new patterns of failures.
\end{itemize}

Finally, to evaluate the difference between "green" vs. "no green" for each user command, LLM, and temperature, we matched the data and computed the following metrics:
\begin{itemize}
    \item \textbf{Boolean difference} is calculated by computing the difference between the \emph{JSON validity} of the "green" prompt and that of the "no green" one.
    \item \textbf{Latency difference} is calculated by computing the difference between the \emph{Latency} of the "green" prompt and that of the "no green" one.
    \item \textbf{Similarity} is computed using Semantic Textual Similarity of BERT over the HomeAssistant JSON representing the routines.
    \item \textbf{Sustainability difference.} The authors compared and analyzed the routine JSONs ("green" vs. "no green"), reporting the patterns in the difference among them.
\end{itemize}

Finally, as reported, for "other" classification in \emph{JSON validity} and \emph{Sustainability difference}, notes were taken. Then, Thematic Analysis~\cite{braun2012thematic} was applied to extract possible common patterns~\cite{braun2021can} and define new failure classes in the former while highlighting possible common patterns and speculating on energy impact in the latter.

\subsubsection{Procedure}
We used a custom code calling LLM endpoints and sent the prompt described above. For each response to the prompts, the code extracts the home automation JSON and sends it to a local instance of the HomeAssistant framework. The script saves locally in CSV files input and output of LLMs and HomeAssistant. Table \ref{tab:analysis1example} reports the head rows of the dataset generated by the process.

Then, metrics defined in Section \ref{sec:home-mockup|subsec:metod-metrics} were computed. Some of them were automatically computed using a Python script while an in-depth analysis by the authors (and Thematic Analysis~\cite{braun2012thematic}) was used to compute others.
Finally, statistical analyses were performed using JAMOVI software~ \cite{csahin2019jamovi}.

All the scripts and datasets mentioned in this Section are publicly available\footnote{\url{https://gitlab.com/i3lab/ecomate}}
and enable full study repeatability.

\begin{table}[ht]
\centering
\caption{Head of the dataset created by the custom code used to generate the comparison between the LLM models.}
\label{tab:analysis1example}

\resizebox{\linewidth}{!}{
\begin{tabular}{ccccccc}
\toprule
    & \textbf{User Command} & \textbf{Type} & \textbf{Category} & \textbf{LLM} & \textbf{Prompt} & \textbf{Temperature} \\
\midrule
0 & make it less chilly in here & immediate & Ambient Temperature & GPT3.5 & No green & 0 \\
1 & make it less chilly in here & immediate & Ambient Temperature & GPT3.5 & Green & 0 \\
2 & make it less chilly in here & immediate & Ambient Temperature & GPT3.5 & No green & 0.7 \\
3 & make it less chilly in here & immediate & Ambient Temperature & GPT3.5 & Green & 0.7 \\
4 & help me cool off & immediate & Ambient Temperature & GPT3.5 & No green & 0 \\
\bottomrule
\end{tabular}
}

\hfill
\vspace{1em}

\resizebox{\linewidth}{!}{
\begin{tabular}{ccccc}
\toprule
    & \textbf{Output} & \textbf{JSON} & \textbf{Latency} & \textbf{JSON validity} \\
\midrule
0 & \textit{Omitted due to excessive length} & \textit{Omitted due to excessive length} & 4400 & False \\
1 & \textit{Omitted due to excessive length} & \textit{Omitted due to excessive length} & 4523 & False \\
2 & \textit{Omitted due to excessive length} & \textit{Omitted due to excessive length} & 4219 & True \\
3 & \textit{Omitted due to excessive length} & \textit{Omitted due to excessive length} & 4849 & True \\
4 & \textit{Omitted due to excessive length} & \textit{Omitted due to excessive length} & 5356 & False \\
\bottomrule
\end{tabular}
}

\hfill
\vspace{1em}

\resizebox{\linewidth}{!}{
\begin{tabular}{cc}
\toprule
    & \textbf{HomeAssistant Response} \\
\midrule
0 & Error while parsing the automation: SyntaxError: Unexpected token c in JSON at position 0 \\
1 & Message malformed: extra keys not allowed @ data['below'] \\
2 & Home-assistant uploaded the automation correctly \\
3 & Home-assistant uploaded the automation correctly \\
4 & Error while parsing the automation: SyntaxError: Unexpected token ; in JSON at position 61 \\
\bottomrule
\end{tabular}
}

\end{table}

\section{Comparative Study  Evaluation}
\label{sec:llm-comparation}

This section aims to report and discuss the results of our empirical evaluation, which investigates the capabilities of different LLMs in generating home automation for controlling smart home environments using the HomeAssistant framework. 
The section is organized as follows. Section \ref{sec:llm-comparation|sub:comparison} reports and discusses the comparison of the general home automation generation capabilities of LLMs, while Section \ref{sec:llm-comparation|sub:comparison-green} focuses on a comparison between a more generic prompt and a one focused on environmental sustainability.

\subsection{LLM as HomeAssistant Controller}
\label{sec:llm-comparation|sub:comparison}
\subsubsection{Results}
Table \ref{tab:recap-metrics} reports the quantitative results linked to our dataset and grouped by LLM, prompt, and temperature used.

\begin{figure}[ht]
     \centering
     \includegraphics[width=\linewidth]{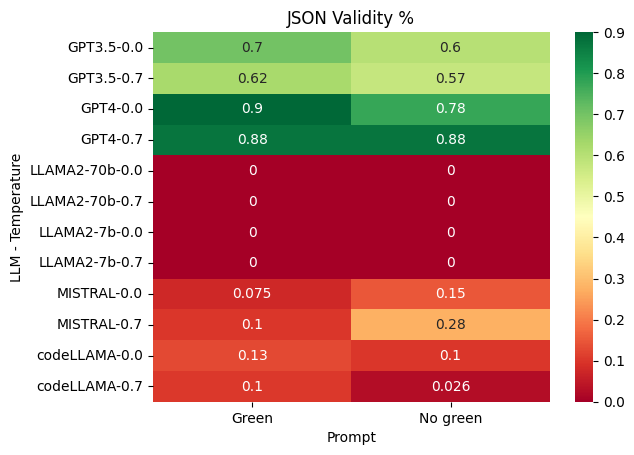}
     \caption{Heatmap of the \emph{JSON validity} percentage for each LLM, prompt and temperature}
     \label{fig:heatmap-comp}
     \Description{Heatmap of the \emph{JSON validity} percentage for each LLM, prompt and temperature}
\end{figure}

Analyzing the results obtained for the \emph{JSON validity} metric, it appears that GPT4 is the best model to generate acceptable routines by HomeAssistant for each prompt and temperature tested. Its predecessor, GPT3.5, also maintained a good \emph{JSON validity} (in the tested cases), while MISTRAL and codeLLAMA have very low scores.
Finally, it is important to highlight that the two tested LLAMA2 models (7b and 70B) obtained the lowest \emph{JSON validity} and did not generate any valid HomeAssistant response.
Additionally, we found that such a difference was statistically significant while considering the LLM, F(5)=209.2066, p$<$.001. There were no significant differences for all the other factors.
Figure \ref{fig:heatmap-comp} shows the heatmap of the \emph{JSON validity} percentage.

In terms of \emph{Latency} to generate HomeAssistant routines, there is a statistically significant difference (F(5)=84.1159, p$<$.001), pinpointing GPT3.5 as the fastest model. Still, we report a low significant effect for the \emph{Relevance score} registered for each LLM model (F(5)=4.06366, p=.007).
A comprehensive overview of the tabular results of the above statistics is reported in \ref{app:comparison-result}.

Finally, the classification of \emph{JSON failure type} found most of the errors in the "Other" group. Still, the second group, the most frequent cause of failure, is "Device → Extra", followed by "Sensor → Trigger value", "Device → Option exists", "Device → Setting", and "Device → Hallucinated". Finally, "Sensor → Suboptimal choice" and "Device → No option exists".

\begin{table}
\centering
\caption{Summary of the metrics results from our empirical evaluation.}
\label{tab:recap-metrics}
\resizebox{\linewidth}{!}{
\begin{tabular}{lll||llll|lll|l} 
\toprule
\multicolumn{3}{l||}{\textbf{Configuration}} & \multicolumn{4}{c|}{\textbf{Relevance}} & \multicolumn{3}{c|}{\textbf{Latency (s)}} & \multicolumn{1}{c}{\multirow{2}{*}{\begin{tabular}[c]{@{}c@{}}\textbf{JSON}\\\textbf{validity (\%)}\end{tabular}}} \\ 
\cline{1-10}
\textbf{LLM} & \textbf{Prompt} & \textbf{t} & \multicolumn{1}{c}{\textbf{Acc}} & \multicolumn{1}{c}{\textbf{FP}} & \multicolumn{1}{c}{\textbf{FN}} & \multicolumn{1}{c|}{\textbf{Rel}} & \multicolumn{1}{c}{\textbf{Min}} & \multicolumn{1}{c}{\textbf{Max}} & \multicolumn{1}{c|}{\textbf{Mean}} & \multicolumn{1}{c}{} \\ 
\hline
\multirow{4}{*}{GPT3.5} & \multirow{2}{*}{Green} & 0 & 0.83 & 0.18 & 0.00 & 0.32 & 2472 & 8955 & 5305.43 & 0.70 \\
    &  & 0.7 & 0.78 & 0.23 & 0.00 & 0.37 & 1704 & 11351 & 4809.83 & 0.63 \\
    & \multirow{2}{*}{No green} & 0 & 0.83 & 0.18 & 0.00 & 0.47 & 2462 & 8907 & 5092.35 & 0.60 \\
    &  & 0.7 & 0.85 & 0.15 & 0.00 & 0.33 & 1990 & 9559 & 5059.35 & 0.58 \\
    \hline
\multirow{4}{*}{GPT4} & \multirow{2}{*}{Green} & 0 & 0.80 & 0.20 & 0.00 & 0.51 & 6926 & 27204 & 15346.60 & 0.90 \\
    &  & 0.7 & 0.80 & 0.18 & 0.03 & 0.52 & 7508 & 30496 & 16481.50 & 0.88 \\
    & \multirow{2}{*}{No green} & 0 & 0.85 & 0.15 & 0.00 & 0.53 & 5996 & 38401 & 16357.30 & 0.78 \\
    &  & 0.7 & 0.83 & 0.18 & 0.00 & 0.54 & 6036 & 22995 & 15458.00 & 0.88 \\
    \hline
\multirow{4}{*}{LLAMA2-70b} & \multirow{2}{*}{Green} & 0 & 1.00 & 0.00 & 0.00 & 0.00 & 10738 & 89666 & 24322.66 & 0.00 \\
    &  & 0.7 & 1.00 & 0.00 & 0.00 & 0.00 & 11259 & 89994 & 24783.03 & 0.00 \\
    & \multirow{2}{*}{No green} & 0 & 1.00 & 0.00 & 0.00 & 0.00 & 9756 & 75800 & 24023.97 & 0.00 \\
    &  & 0.7 & 1.00 & 0.00 & 0.00 & 0.00 & 7196 & 64823 & 20742.31 & 0.00 \\
    \hline
\multirow{4}{*}{LLAMA2-7b} & \multirow{2}{*}{Green} & 0 & 1.00 & 0.00 & 0.00 & 0.00 & 5392 & 41343 & 12942.93 & 0.00 \\
    &  & 0.7 & 1.00 & 0.00 & 0.00 & 0.00 & 5190 & 34398 & 13133.83 & 0.00 \\
    & \multirow{2}{*}{No green} & 0 & 1.00 & 0.00 & 0.00 & 0.00 & 6000 & 41947 & 12981.10 & 0.00 \\
    &  & 0.7 & 1.00 & 0.00 & 0.00 & 0.00 & 3332 & 42804 & 14032.03 & 0.00 \\
    \hline
\multirow{4}{*}{MISTRAL} & \multirow{2}{*}{Green} & 0 & 0.95 & 0.00 & 0.05 & 0.58 & 963 & 18851 & 9915.65 & 0.08 \\
    &  & 0.7 & 0.98 & 0.00 & 0.03 & 0.41 & 1065 & 29623 & 10636.85 & 0.10 \\
    & \multirow{2}{*}{No green} & 0 & 0.98 & 0.00 & 0.03 & 0.54 & 1110 & 28178 & 10605.33 & 0.15 \\
    &  & 0.7 & 0.95 & 0.03 & 0.03 & 0.58 & 912 & 33440 & 9930.43 & 0.28 \\
    \hline
\multirow{4}{*}{codeLLAMA} & \multirow{2}{*}{Green} & 0 & 0.95 & 0.05 & 0.00 & 0.24 & 9383 & 34592 & 16133.15 & 0.13 \\
    &  & 0.7 & 0.87 & 0.10 & 0.03 & 0.40 & 5262 & 35300 & 13351.74 & 0.10 \\
    & \multirow{2}{*}{No green} & 0 & 0.95 & 0.05 & 0.00 & 0.38 & 9193 & 46016 & 17245.23 & 0.10 \\
    &  & 0.7 & 0.97 & 0.03 & 0.00 & 0.29 & 8543 & 46445 & 14230.82 & 0.03 \\
\bottomrule
\end{tabular}
}
\end{table}

\subsubsection{Discussion}

\paragraph{Overall LLMs' Comparison}
The most evident and promising result is that GPT models are able to correctly generate HomeAssistant routines in most of the cases under evaluation, while MISTRAL and LLAMA are not.

It is important to emphasize that both quantitative and an extensive qualitative analysis of JSON generation found that such output is generally on point, meaning that each LLM is able to understand the user command and consider (as well as activate) in the home automation routine the devices needed to achieve the goal.
The main reason for the failure of home automation routines is the malformation of the message sent to HomeAssistant, which is evident in LLAMA and MISTRAL.

In particular, LLAMA doesn't show any knowledge about the JSON format of HomeAssistant API since there is no apparent difference in the outputs generated by any of the LLAMA models (i.e., 7B and 70B), leading to very negative results.
A small improvement is shown by the codeLLAMA model, which is a more specific version for code-related tasks, such as understanding and code generation since an apparent structure can be identified in the complements of the model. However, this improvement does not result in the correctness of the routines accepted by HomeAssistant, as the number of valid JSONs remains very low. MISTRAL yields a slightly better result than LLAMA. However, the format issue still applies. This shows a clear need for fine-tuned models or the employment of prompt engineering techniques, such as giving examples (few or N shot learning) and clear instructions on the HomeAssistant JSON format. For the latter proposed technique, it is worth considering the increase in tokens needed for each message.

\paragraph{Message malformation}
The analysis of the malformatted messages found issues with the keywords used by the models. For instance, the name of the routine needs to be specified under the "alias" keyword, as correctly set by GPT models. However, it does not translate to other models, as most of the time, the JSON presents the word \textit{"name"} (particularly in LLAMA models) or \textit{"algorithm"} (for MISTRAL), raising the principal cause of failure.

Similarly, instead of \textit{"trigger"}, \textit{"triggers"} (plural form) or \textit{"triggs"} are often used by LLAMA or MISTRAL.
Another evident instance of message malformation is the presence of Python code in the output - either completely or in some parts of the routine. Indeed, the routine randomly calls a made-up script or hallucinates some non-existent functions to trigger the requested command, e.g., the use of \texttt{stop\_cleaning()} for user commands that fall in the "Robot Control" category.

\paragraph{Relevant devices}
One of the metrics targeted by the study is \emph{Relevance score}, defined as the ability of LLMs to employ the relevant devices needed to reach the goal of the user command. Each user command is grouped in one of the seven categories. Each category is associated with a set of appliances relevant to the specific objective. For instance, in a "Robot Control" command, the vacuum cleaner is expected to be activated/deactivated, while the coffee machine is not. The study revealed that all the generative models analyzed yielded good results, as in the majority of the cases, all the relevant devices were targeted.
Still, GPT models were found to be quite good, particularly GPT4, as they are extremely precise in the choice of the relevant appliances for the HomeAssistant automation.

However, it is worth mentioning that, in most cases, extra devices were targeted, meaning that in addition to the relevant ones, other unnecessary appliances were used in the routine generation process. Such difference is evident (even if not statistically significant) in the "green" prompt, where the presence of instructions to be more energy efficient was translated into a general habit of turning off all the appliances of the house, or at least a good amount of them.
In particular, two command categories presented a decrease in reliability than others: "Security" and "Other Appliances". For instance, if the LLM is asked to clean the house, the generated routine contains the script to activate the vacuum cleaner; however, it could also randomly contain turning off all the lights, the security camera, or other appliances present in the house. 

\paragraph{Triggers} 
Another cause of failure is the trigger values. Logically speaking, the triggers were correctly managed, but the use of too general metrics caused a failure in the routine import process. When asked to use natural light instead of light bulbs, the general behavior of the models was to set as the starting point of the routine the sunrise; similarly, when asking to make the house more comfortable at night, the time of the routine is "sunset". This practice raises an error as HomeAssistant asks for a specific timestamp format. Similarly, qualitative analysis shows that, in the "Ambient Temperature" commands, sometimes the routine is triggered when a "hot" or "cold" temperature is reached instead of setting the precise amount in degrees. The generative ability of LLM models is to grasp when it is right to activate a routine following the user's instruction, but still, there is not a very precise knowledge translation in HomeAssistant routine formatting.
Finally, the proposed temperatures and hours were coherent with the task; for example, night-related commands were generated with timestamps at night by all LLMs. However, the range vastly varied between answers, both for timestamps and temperatures. In the latter example, indeed, a cold temperature can be interpreted as 16°C or 20°C.

\paragraph{Hallucinations}
In some rare cases, the response is completely hallucinated, meaning that very unusual outputs were returned. The main reason for the hallucination is the presence of non-existent devices (like a washing machine, a dryer, a kitchen hob, or a microwave), followed by the prompt leak of the energy consumption. Lastly, very non-green and non-sense actions like turning on all the lights in the house were generated, especially when the command involved a strong natural language (e.g., \emph{"Make it less stuffy in here"}).

\paragraph{General and imprecise commands}
As probably expected, "broad" user commands sometimes create misunderstandings in the generation process. Commands like \emph{"make it less stuffy in here"} or \emph{"help me wind down"} are too general and open to different interpretations to create a very homogeneous response from each LLM. This generalization of the prompt has caused, in some cases, some uncertainty, meaning that the same LLM has generated a very different output from the same input. On the other hand, the study faced some very interesting responses as the generative models were able to associate music with an emotional state. As an example, in response to \textit{"make it cozy in here"}, GPT3.5 suggested the song 'Shape of you' by Ed Sheeran on the media player that is, generally speaking, a quite chilling and welcoming song. Similarly, in other cases, some Spotify tracks were played on the smart speaker (even though some were just random song IDs that didn't exist on the streaming platform). One more thing that is worth mentioning is that GPT3.5 suggested "Never Gonna Give You Up" by Rick Astley as a response to \textit{"Make it cozy in here"}. This shows that GPT has indirect embedded knowledge of the trending internet meme called rick-rolling, which is the process of playing a prank on the user by making him open the link of that particular song. This is probably caused by training data in the model that were taken from shared routine examples in forums or other online repositories.

\begin{figure}[ht]
     \centering
     \begin{subfigure}[b]{0.47\linewidth}
         \centering
         \includegraphics[width=\linewidth]{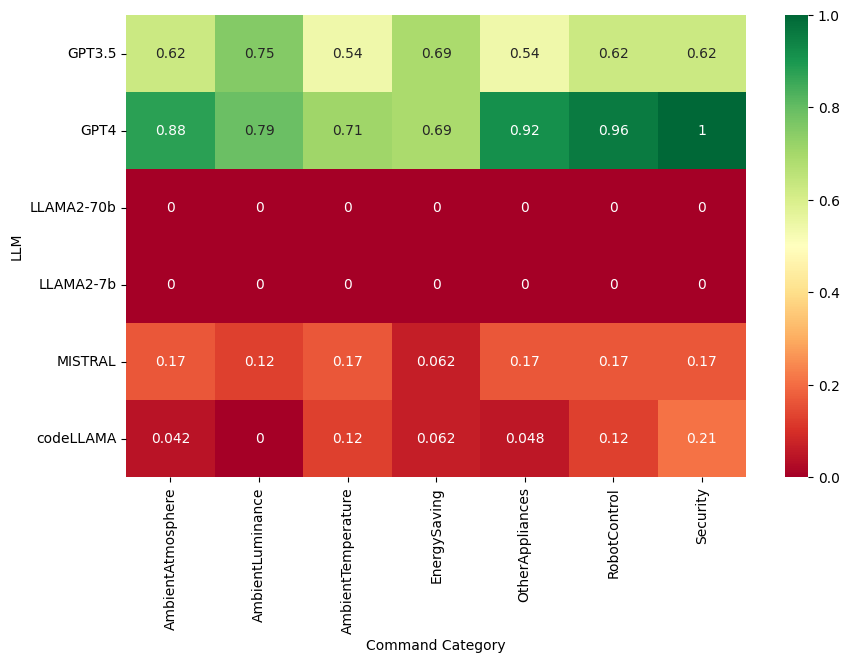}
         \caption{}
     \end{subfigure}
     \begin{subfigure}[b]{0.52\linewidth}
         \centering
         \includegraphics[width=\linewidth]{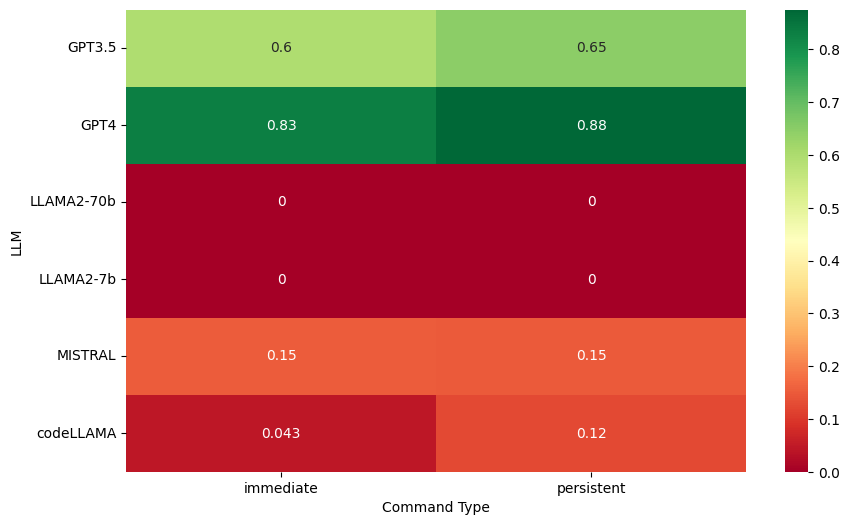}
         \caption{}
     \end{subfigure}
    \caption{Heatmap of the \emph{JSON validity} percentage for each LLM, and Command Category (a) or Command Type (b)}
    \Description{Heatmap of the \emph{JSON validity} percentage for each LLM, and Command Category (a) or Command Type (b)}
    \label{fig:heatmap_by_cmd}
\end{figure}

\paragraph{User Command Type}
The empirical results highlighted that the broad meaning of some user commands misleads the ability of the models to distinguish between an immediate and a persistent command (as also highlighted by Figure \ref{fig:heatmap_by_cmd}). The two different types of inputs were employed to test the capacity of the LLMs to understand the difference between a single action that has to be done once (i.e., immediate command) and a routine that has to be repeated over time (i.e., persistent commands). What emerges is that every command is interpreted as a persistent one, and quite always, a repeating routine is generated. This is due, probably, to the presence of the words 'routine' and 'automation' often repeated in the prompt. Followup research in the field could investigate if clearly distinguishing the difference of user command types in the prompt could lead to better results, as none of the LLMs embed such a concept.

\paragraph{Separation of the code from the natural language}
Even if this happens just a few times in different generations, it should be mentioned that sometimes the output was not properly formatted as asked in the prompt. Only GPT models were able to correctly separate the text response from the JSON code as specified in the instructions in the prompt. MISTRAL and LLAMA randomly inserted in the JSON words like \emph{"code"} or wrongly inserted back-ticks before the routine script, making the extracting and parsing process difficult to successfully run the routine on the HomeAssistant framework. 

\subsection{LLM as sustainable controllers}
\label{sec:llm-comparation|sub:comparison-green}

\subsubsection{Results}
Table \ref{tab:recap-metrics-green} reports the quantitative results linked to our dataset divided by the prompts and grouped by LLM and temperature used.

\begin{figure}[ht]
     \centering
     \begin{subfigure}[b]{0.5\linewidth}
         \centering
         \includegraphics[width=\linewidth]{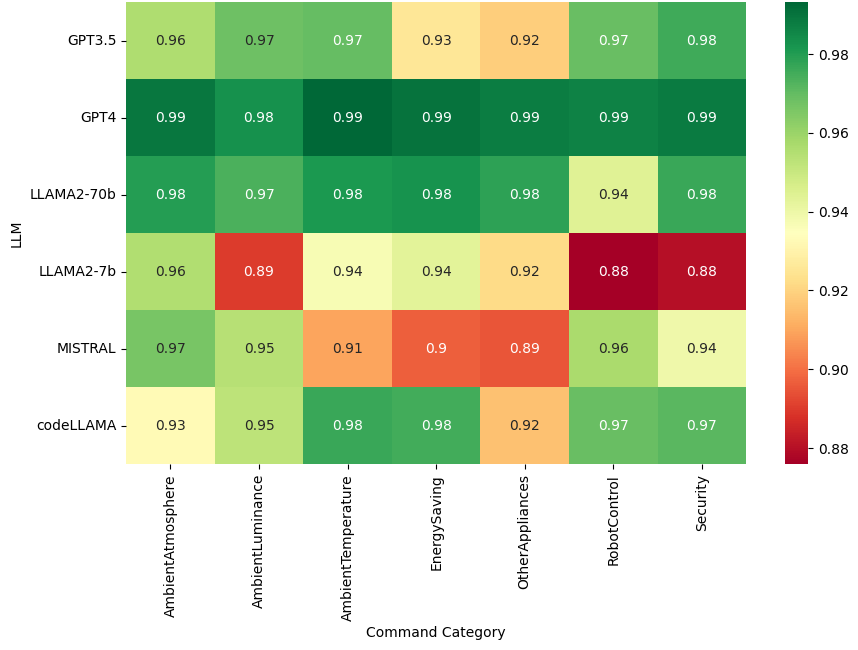}
         \caption{Heatmap of the \emph{Similarity} percentage for each LLM, and user command category}
         \label{fig:heatmap-sim}
         \Description{Heatmap of the \emph{Similarity} percentage for each LLM, and user command category}
     \end{subfigure}
     \begin{subfigure}[b]{0.49\linewidth}
         \centering
         \includegraphics[width=\linewidth]{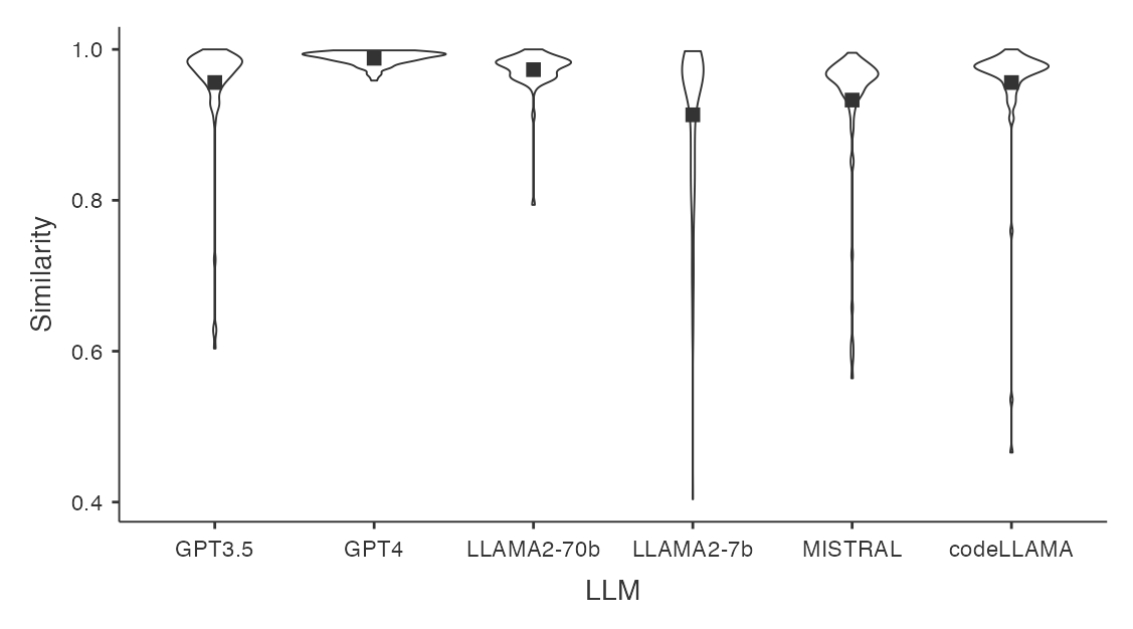}
         \caption{Violin map of the \emph{Similarity} distribution for each LLM}
         \label{fig:similarity_box}
         \Description{Violin map of the \emph{Similarity} distribution for each LLM}
     \end{subfigure}
    \caption{Plots of the \emph{Similarity} metric taken into consideration in the evaluation of LLM capabilities to generate HomeAssistant JSONs}
    \label{fig:sim-figures}
\end{figure}

Analyzing the \emph{Similarity} of the JSON routine for both "green" and "no green" prompts, results (as depicted in Figure \ref{fig:sim-figures}) show that GPT4 has the highest mean similarity, followed by LLAMA2-70b, GPT3.5 and codeLLAMA, MISTRAL, and LLAMA2-70b. Such difference is statistically significant, with F(5)=9.910, p$<$.001.

The computed \emph{Boolean difference} pinpoints that -- except for MISTRAL --  LLM models overall tend to have the same \emph{JSON validity} outcome for both "green" and "no green" prompts.

Finally, GPT4, LLAMA2-70b and MISTRAL produced a positive \emph{Latency difference}, thus highlighting that, on average, the "green" prompts have a higher latency than their "no green" counterparts. In contrast, GPT3.5, LLAMA2-7b, and codeLLAMA had a negative mean \emph{Latency difference}.

As already declared before, a comprehensive overview of the tabular results of the mentioned statistics is reported in  \ref{app:comparison-result}.

\begin{table}[ht]
\centering
\caption{Summary of the metrics results from our empirical evaluation of "green" vs. "no green" prompts.}
\label{tab:recap-metrics-green}
\resizebox{\linewidth}{!}{
\begin{tabular}{lll||rrr}
\toprule
\multicolumn{1}{c}{\textbf{\textit{~}}} & \multicolumn{1}{c}{\textbf{LLM}} & \multicolumn{1}{c||}{\textbf{t}} & \multicolumn{1}{c}{\textbf{Boolean difference}} & \multicolumn{1}{c}{\textbf{Similarity}} & \multicolumn{1}{c}{\textbf{Latency difference}} \\ 
\hline
\multirow{12}{*}{\textit{Mean ~ ~ ~ ~ ~ ~ ~ ~ ~ ~ ~}} & GPT3.5 & 0.0 & 0.100 & 0.968 & 213 \\
    & ~ & 0.7 & 0.0500 & 0.944 & -250 \\
    & GPT4 & 0.0 & 0.125 & 0.989 & -1011 \\
    & ~ & 0.7 & 0.00 & 0.988 & 1024 \\
    & LLAMA2-70b & 0.0 & 0.00 & 0.974 & -1760 \\
    & ~ & 0.7 & 0.00 & 0.973 & 4100 \\
    & LLAMA2-7b & 0.0 & 0.00 & 0.913 & -38.2 \\
    & ~ & 0.7 & 0.00 & 0.914 & -898 \\
    & MISTRAL & 0.0 & -0.0750 & 0.932 & -690 \\
    & ~ & 0.7 & -0.175 & 0.934 & 706 \\
    & codeLLAMA & 0.0 & 0.0256 & 0.974 & -849 \\
    & ~ & 0.7 & 0.0769 & 0.938 & -879 \\ 
\hline
\multirow{12}{*}{\textit{Standard deviation ~ ~ ~ ~ ~ ~ ~ ~ ~ ~ ~}} & GPT3.5 & 0.0 & 0.709 & 0.0618 & 1611 \\
    & ~ & 0.7 & 0.749 & 0.0858 & 2477 \\
    & GPT4 & 0.0 & 0.463 & 0.00911 & 5881 \\
    & ~ & 0.7 & 0.453 & 0.00867 & 5457 \\
    & LLAMA2-70b & 0.0 & 0.00 & 0.0367 & 9537 \\
    & ~ & 0.7 & 0.00 & 0.0159 & 17439 \\
    & LLAMA2-7b & 0.0 & 0.00 & 0.125 & 7969 \\
    & ~ & 0.7 & 0.00 & 0.0880 & 6118 \\
    & MISTRAL & 0.0 & 0.350 & 0.103 & 5209 \\
    & ~ & 0.7 & 0.549 & 0.0828 & 9031 \\
    & codeLLAMA & 0.0 & 0.362 & 0.0193 & 8046 \\
    & ~ & 0.7 & 0.354 & 0.110 & 8702 \\
\bottomrule
\end{tabular}
}
\end{table}

\subsubsection{Discussion}

The examination of \emph{Boolean difference} and \emph{Latency difference} metrics failed to reveal any significant distinctions between prompts labeled as "green" and those labeled as "no green." This lack of divergence was observed across two key dimensions: the number of JSON configurations accepted by HomeAssistant and their generation time.

On the contrary, \emph{Similarity} was found statistically different, and a qualitative exploration of the JSON pairs revealed nuanced insights. Notably, prompts designated as green elicited a greater propensity for generating routines with green attributes. For example, these routines exhibited a higher frequency of actions, such as appliance deactivation, lower temperature settings, and increased trigger section evaluations. Although instances of high similarity between JSON outputs were prevalent, discrepancies primarily manifested in trigger values. Nevertheless, the underlying thematic coherence remained intact, with both sets of prompts predominantly employing the same appliances. In isolated cases, minor deviations were noted, such as the inclusion of additional appliances -- such as vacuum cleaners or coffee machines -- in one of the JSON outputs.

We tried to formally explore the comparison between the amount of energy (in terms of kWH) used by each routine ("green" or "no green" prompts).
However, the nature of the format in which the JSON in which HomeAssistant is formatted did not allow us to easily read and estimate the kWh consumed by this routine.
Therefore, we subsequently proceeded with a more qualitative approach by viewing the different responses in pairs. Even with a qualitative approach, estimating their consumption quantitatively was impossible. 
Regardless, what emerged from the qualitative analysis is that disparities in appliance selection were often attributed to the presence of thermostat management within green prompts, a feature conspicuously absent in no-green prompts. Additionally, green prompts tended to induce a higher incidence of hallucinated appliances compared to their no-green counterparts.
For example, if the LLM is asked to clean the home, the "green" routine will include the script to operate the vacuum cleaner, but it may also include arbitrarily turning off all of the lights, the security camera, and other appliances in the house.
In the future, to overcome the above-mentioned limitation of the kWh estimation with the two different types of routines ("green" or "no green" prompts), an initial analysis involving a smaller number of appliances and a simpler house layout could be carried out and feasible to provide a framework to deliver such estimation.

According to \citet{rillig2023risks}, LLMs and other new innovative technologies (e.g., Metaverse) directly and indirectly impact the environment with also implications on environmental research. One of the direct negative impacts is the high amount of energy consumption LLMs need to be trained and employed. In particular, \citet{luccioni2023estimating} quantified the footprint of BLOOM 176B parameters LLM, considering the training equipment manufacturing, training the model, and model deployment (accessible via API endpoints). In the same way,  \citet{faiz2023llmcarbon} proposed a framework to compute the carbon footprint of different AI models and estimate the carbon emitted across usage.
However, \citet{tomlinson2024carbon} reported comparing the usage of generative AI systems and human individuals performing equivalent writing and illustrating tasks. Contrary to expectations, AI systems released between 130 and 1500 times less carbon per page of text generated than their human counterparts, and similar results (310-2900 times less) were found for image generation. The authors discussed how generative AI is not a replacement for human tasks; however, it holds the potential to perform some activities with much lower carbon emissions.
In our specific research context, it was not possible to compare or compute watt differences to evaluate whether LLMs can provide home automation that can significantly reduce energy consumption and positively impact carbon emissions, also considering in the formula the footprint of LLM in the overall energy balance.

\subsection{Key Findings}
The empirical evaluation of different LLM-based HomeAssistant controllers reveals several key findings:

\begin{itemize}
    \item \textbf{Performance Variation.} GPT models demonstrate superior performance in generating HomeAssistant routines compared to MISTRAL and LLAMA variants. GPT4, in particular, exhibits the highest \emph{JSON validity}, while GPT3.5 is the fastest model. LLAMA models, especially LLAMA2, struggle with \emph{JSON validity} due to a lack of knowledge about the HomeAssistant API format.
    \item \textbf{Message Malformation.} Issues with keyword usage and trigger values contribute to routine failure. LLAMA and MISTRAL often produce malformatted JSON, hindering routine execution. Common issues include incorrect keywords like "name" instead of "alias" and imprecise trigger values.
    \item \textbf{Relevance and Hallucinations.} Despite generally selecting relevant devices, LLMs often include unnecessary appliances in routines, especially in response to broad commands. Hallucinated responses occasionally occur, with non-existent devices or irrelevant actions generated.
    \item \textbf{User Command Interpretation.} LLMs struggle to differentiate between immediate and persistent commands, often generating routines as if all commands were recurring tasks. Clearer prompts may improve model performance in distinguishing command types.
    \item \textbf{JSON Formatting.} GPT models excel in separating code from natural language, ensuring proper JSON formatting as specified in the prompt. However, LLAMA and MISTRAL occasionally insert incorrect code elements, complicating routine execution.
    \item \textbf{Impact of "Green" Prompts.} While quantitative metrics like \emph{JSON validity} show no significant difference between "green" and "no green" prompts, qualitative analysis reveals nuanced distinctions. Green prompts elicit a greater propensity for generating routines with green attributes, such as appliance deactivation and lower temperature settings. Discrepancies primarily manifest in trigger values, with green prompts inducing a higher incidence of hallucinated appliances compared to no-green prompts.
\end{itemize}

Overall, while GPT models demonstrate strong performance in generating HomeAssistant routines, challenges remain in ensuring accurate JSON formatting and interpreting user commands effectively, with green prompts showing subtle yet impactful differences in routine generation. Further research could explore prompt engineering or fine-tuning techniques to enhance model understanding and generate more precise routines, particularly in sustainability-focused contexts.

\section{The \systemName{} Design}
\label{sec:ecomate-design}

\systemName{} is a web-based application that enables interaction with an LLM-based chatbot to create HomeAssistant routines and manage the automation of the house.
The main features of the application are:
\begin{itemize}
    \item \textbf{Green energy orientation}. Guide users towards more sustainable energy usage. 
    \item \textbf{Smart routine automation}. Enable users to automate their home devices in an energy-efficient manner.
    \item \textbf{Seamless integration with HomeAssistant}. Ensure compatibility with existing commercial home automation environments.    
\end{itemize}

The section is organized as follows, aiming to organize a complete system description. Section \ref{sec:ecomate-design|subsec:system-ui} describes the design of the User Experience of the entire \systemName{} application. Section \ref{sec:ecomate-design|subsec:system-implementation} details the implementation, while Section \ref{sec:ecomate-design|subsec:system-gpt} presents the integration of the GPT model.

\subsection{User Experience}
\label{sec:ecomate-design|subsec:system-ui}
The design philosophy is rooted in two fundamental aspects: the system's simplicity and a focus on friendly and easy interaction with the home automation system, emphasizing a conversational approach reified in the \systemName{} chatbot.

The application core lies in the \emph{AI Chat Helper} page, allowing users to pose questions regarding energy consumption, smart automation, and environmentally friendly practices. The interface adopts a chat-based format, presenting messages in bubbles as in traditional chat UX. Noteworthy UI buttons enable the facility to generate, save, submit, and inspect routines produced by the LLM. Additionally, a small banner suggests conversation "starters" to users, possibly enhancing the engagement and usability of the application.

\begin{figure}[ht]
    \centering
    \includegraphics[width=0.75\linewidth]{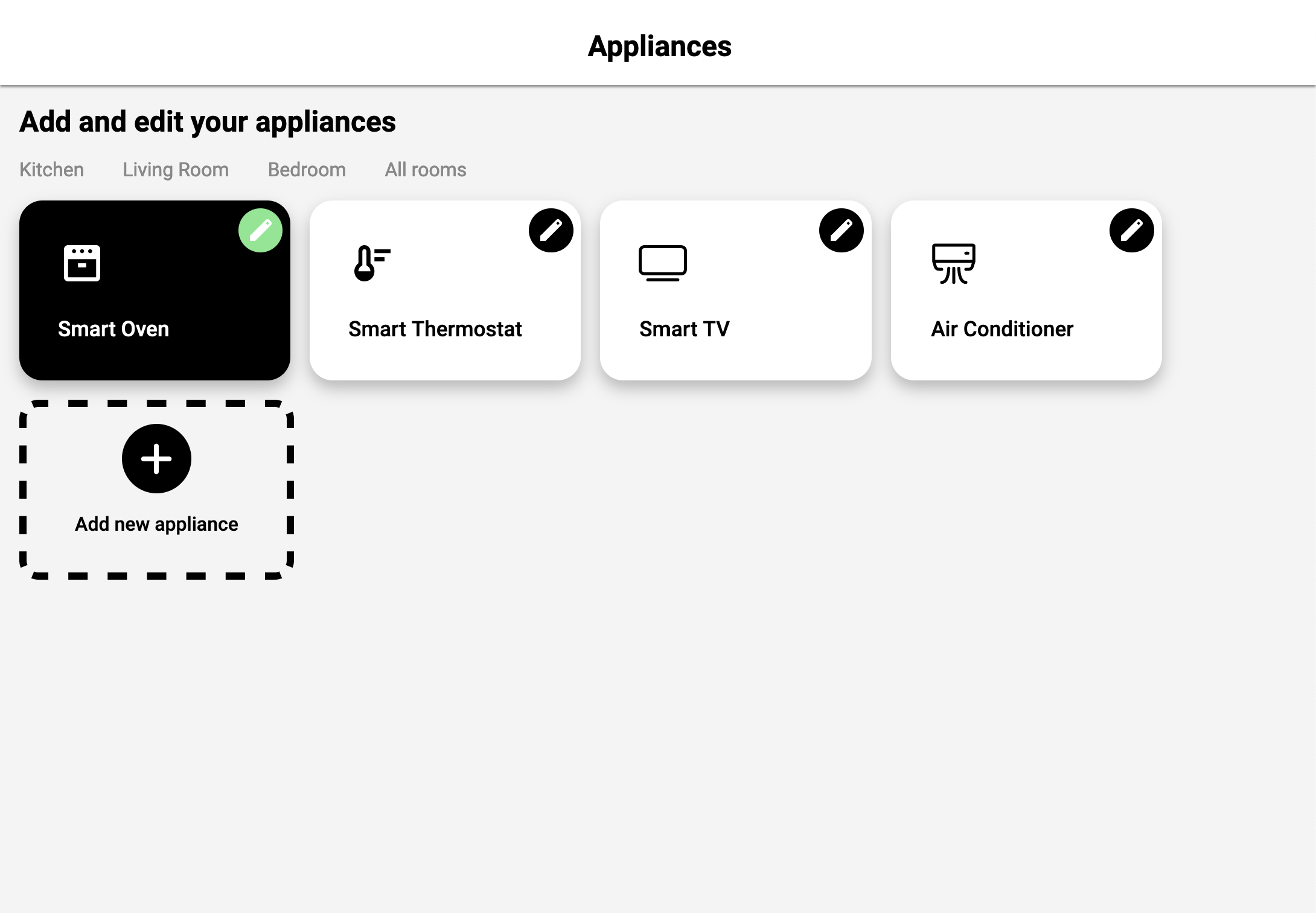}
    \caption{Snapshot of the appliances' page}
    \label{figure:appliances-page}
    \Description{Snapshot of the appliances' page}
\end{figure}

In addition, a dedicated page for managing \emph{Appliances} facilitates a deeper understanding of energy usage. A list of home devices, filterable by room, is prominently displayed. Users can edit existing appliances and seamlessly introduce new ones through a user-friendly combination of buttons and forms (see Figure \ref{figure:appliances-page}).

The \emph{Routines} page allows users to manage the code associated with routines generated by the chatbot. Displayed in a list format, routines are accompanied by interactive buttons for inspection and deletion, giving users control over their automated processes.

Finally, a \emph{Settings} page offers a straightforward and user-friendly form for configuring system integration with HomeAssistant.

\subsection{Integration of GPT with HomeAssistant}
\label{sec:ecomate-design|subsec:system-gpt}

The GPT3.5 plays a central role in \systemName{} application since it allows unlocking a conversational interaction with users and generates home automation routines.
As discussed in the previous evaluation (Section \ref{sec:home-mockup|subsec:prompt}), a prompt engineering technique was applied to generate meaningful and sustainable home automation.

The prompt sent to the GPT3.5 has an environmentally sustainable flavor and follows the structure previously defined. However, unlike the previous setup, the LLM is interposed using a small subset of the chat history (last 5 messages), and the smart home configuration (appliances' location, type, and consumption) that is slightly different from the previous one but allows a more resilience to the conversation context (thanks to the chat history) and aims to be more user-friendly.
A complete overview of the prompt is provided in \ref{app:specific-design-promt}.

\begin{figure}[ht]
    \centering
    \includegraphics[width=0.75\linewidth]{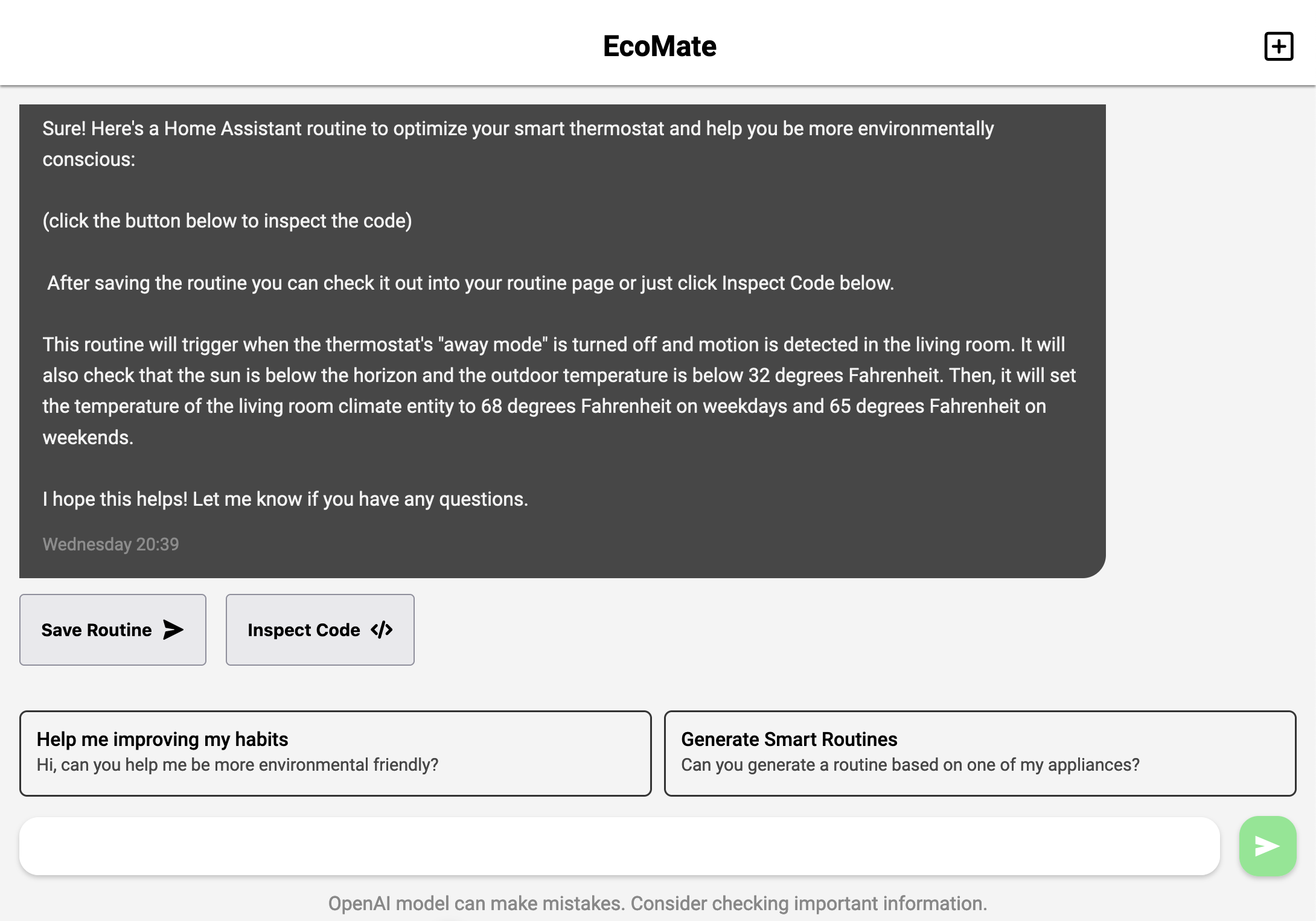}
    \caption{Example of a chat generating a home automation}
    \label{figure:routine_chat}
    \Description{Snapshot of a chat response in which home automation is generated}
\end{figure}

As highlighted before, using such prompt engineering technique, GPT3.5 usually generates a response including a JSON structure, parsable by HomeAssistant's API. The application logic locates the JSON and automatically integrates the routine into the home automation system. In addition, the JSON is removed from the chat to make the answers more user-friendly (as in the example of a chat shown by Figure \ref{figure:routine_chat}), and two buttons "Save Routine" and "Inspect Code" are added in the system UI right after the chat message.

\subsection{Implementation Details}
\label{sec:ecomate-design|subsec:system-implementation}

The software was designed according to a standard \emph{three-tier client-server architecture}, enabling the split of the logic part from a graphical interface. The system is divided into a backend implemented in \emph{Node.js}, and a frontend in \emph{Vue.js}.
The backend also relies on an external authentication system and database implemented with \emph{Supabase}. In addition, community packages to connect to OpenAI services and the HomeAssistant framework are used.
All the software is being released as open source\footnote{\url{https://gitlab.com/i3lab/ecomate}}.

\section{\systemName{} User Evaluation}
\label{sec:ecomate-user-eval}

The second empirical study aims to evaluate \systemName{} application and addresses the third research question (i.e., RQ3 as defined in Section \ref{sec:intro}). It touches two main areas: \emph{likability and dialogue}, which refers to how much the chatbot is appreciated and how smooth the dialogue with it, and \emph{usability}, which is the grade to which participants are able to utilize the chatbot.

\emph{likability} refers to how much a tool is appreciated and how simple it is to use; \emph{usability} is the degree to which something can be utilized, including an assessment of its interaction paradigms.

Notably, the first evaluation presented in this paper (Section \ref{sec:llm-comparation}) showed that GPT4 is the best LLM model for the specific task under study. However, for ethical reasons, at the time of the study (i.e., early 2024), it was not possible to create an instance of GPT4 on Azure OpenAI Service and thus preserve user data within our university organization without it being used by a third party (i.e., OpenAI). Therefore, for this study, we used the second best (and the ethically compliant) solution, which is GPT3.5 via Azure OpenAI Service. 



\subsection{Methodology}

A between-subject design was used for the experiment, with \textit{Group} as the fixed factor. Participants in the \textit{RULE} condition interacted with a traditional rule-based chatbot to create home automation, while in the \textit{LLM} condition chatted with the LLM agent.

\subsubsection{Research Variables}
\label{sec:ecomate-user-eval|subsec:study-researchvar}
A digital survey was used to collect data. In response to questions, utilizing a 7-point Likert scale that runs from 1 ("Completely Disagree") to 5 ("Completely Agree"),participants reported the following quantitative measures:
\begin{itemize}
    \item \textit{Parasocial Interaction (PSI) scale}~\cite{horton1956mass, tsai2021chatbots} ($\alpha$ = 0.86) assessed the extent to which participants felt a sense of attachment, connection, and dialogue capabilities with the chatbot;
    \item \textit{System Usability Scale (SUS)}~\cite{brooke1996sus, bangor2009determining} ($\alpha$ = 0.68) was employed to ascertain how participants judged the usability of the chatbot.
\end{itemize}

\subsubsection{Participants}
The study involved 56 participants with a median age of 24 years (range 21-30, M=24.6, SD=2.21); a more granular view of participants' demographics is reported in Table \ref{table:age}. Using snowball sampling to gather participants (initiated by our friends, coworkers, and other master students), all persons gave their consent to participate, signing a consent form informing them about procedures, goals, and data treatment. The study was conducted in compliance with the Helsinki Declaration of 1975 (as revised in 2008) and received ethical approval from the Ethical Committee of our University. The study was conducted using the same laptops in our research laboratory.

\begin{table}[ht]
\caption{Participants age demographics detail}
\label{table:age}
\begin{tabular}{llrrr} 
\toprule
\multicolumn{1}{c}{\textbf{Group}} & \multicolumn{1}{c}{\textbf{Gender}} & \multicolumn{1}{c}{\textbf{N}} & \multicolumn{1}{c}{\textbf{Mean}} & \multicolumn{1}{c}{\textbf{SD}} \\ 
\midrule
LLM & Female & 15 & 25.9 & 2.75 \\
~ & Male & 15 & 23.5 & 1.68 \\
RULE & Female & 6 & 24.5 & 1.87 \\
~ & Male & 20 & 24.5 & 1.79 \\
\bottomrule
\end{tabular}
\end{table}

\subsubsection{Procedure}
Three stages make up the experimental procedure, and each phase lasts around fifteen minutes in total.
In the first step, participants were invited to fill out a generic biographical questionnaire. A researcher also used an additional paper page to present the study's tasks and situation (available in \ref{app:scenario-detail}), equal for both conditions.
Participants were randomly assigned to one of the between-study conditions and asked to engage with the chatbot and complete the tasks during the second step of the experiment.
At the end of the study, participants completed a questionnaire that included all of the questions needed to evaluate the research variables that were discussed in Section \ref{sec:ecomate-user-eval|subsec:study-researchvar}, fully clarified in \ref{app:questionnaire}.

\subsection{Results and Discussion}

\subsubsection{Results}
In the LLM condition, participants reported in the \textit{Perceived Parasocial Interaction (PPI)} of the \textit{PSI scale} a mean of 4.90 (SD=1.10), while its \textit{Perceived Dialogue (PD)} was an average of 5.79 (SD=0.75), and \textit{Interaction Satisfaction} averaged 6.33 (SD=0.45). The SUS score has a mean of 86.49 (SD=7.84).
While, in the RULE condition, participants reported in the \textit{Perceived Parasocial Interaction (PPI)} an average of 3.67 (SD=0.94), while its \textit{Perceived Dialogue (PD)} had a mean of 5.07 (SD=0.92), and \textit{Interaction Satisfaction} averaged 3.04 (SD=0.37). The SUS score has a mean of 83.57 (SD=10.47).
Table \ref{table:desc_results} structures the above-described results.

\begin{table}[ht]
\centering
\caption{Descriptive Results of the Research Variables}
\Description{The table shows the descriptive results (average and standard deviation) of the three scales evaluated}
\label{table:desc_results}
\begin{tabular}{lllrr} 
\toprule
 & \multicolumn{1}{c}{~} & \multicolumn{1}{c}{\textbf{Group}} & \multicolumn{1}{c}{\textbf{AVG}} & \multicolumn{1}{c}{\textbf{SD}} \\
\midrule
\multirow{6}{*}{\textit{PSI scale}} & PPI & LLM & 4.90 & 1.10 \\
 & ~ & RULE & 3.67 & 0.94 \\
 & PD & LLM & 5.79 & 0.75 \\
 & ~ & RULE & 5.07 & 0.92 \\
 & IS & LLM & 6.33 & 0.45 \\
 & ~ & RULE & 3.04 & 0.37 \\
\hline
\multirow{2}{*}{\textit{SUS scale}} &  & LLM & 86.49 & 7.84 \\
 &  & RULE & 83.57 & 10.47 \\
\bottomrule
\end{tabular}
\end{table}

An independent samples t-test was performed to evaluate the difference between the PSI and SUS scales' scores between the two groups -- LLM (30 participants) and RULE (26 people).
While, the SUS did not show any statistical difference t(54)=1.19, p=0.120, all the measures in the PSI scale are statistically different. In particular, the PPI is t(54)=4.46 with p$<$.001, the PD is t(54)=3.26 p$<$.001, and IS is t(54)=29.67 p$<$.001.
As before, Table \ref{table:independent} structures the independent samples t-test results.

\begin{table}[ht]
\centering
\caption{Independent Samples T-Test result}
\label{table:independent}
\begin{tabular}{llrrr}
\toprule
\multicolumn{1}{c}{~} & \multicolumn{1}{c}{~} & \multicolumn{1}{c}{\textbf{Statistic}} & \multicolumn{1}{c}{\textbf{df}} & \multicolumn{1}{c}{\textbf{p}} \\
\midrule
PPI & Student's t & 4.46 & 54.0 & $<$.001 \\
PD & Student's t & 3.26 & 54.0 & $<$.001 \\
IS & Student's t & 29.67 & 54.0 & $<$.001 \\
SUS & Student's t & 1.19 & 54.0 & 0.120 \\
\bottomrule
\end{tabular}

\vspace{1ex}
{Note. $\mathit{H_{a}}:\; \mathit{\mu_{LLM}} > \mathit{\mu_{RULE}}$}
\end{table}

\subsubsection{Discussion}

As stated before, this second study aimed to assess the likability and usability of an LLM-based chatbot versus its traditional rule-based counterpart in promoting more sustainable home automation routines.
Regarding usability, according to \citet{bangor2009determining}, both the chatbots have usability more than acceptable (graded with B), which implies an adjective for the system that is between "excellent" and "best imaginable". In addition, since no statistical difference was found, participants found both chatbots usable.

Conversely, the results on the Parasocial Interaction scale have a statistically significant difference, implying that the LLM-based version outperformed the rule-based one, indicating a pleasant and successful interaction between participants and \systemName{}.
In addition, the results on the PSI scale are higher than those previously found by \citet{tsai2021chatbots}. Specifically, users reported increased levels of IS and PPI.

Using an LLM instead of a standard rule-based chatbot resulted in greater outcomes \cite{raiaan2024review} since LLMs have enhanced language capabilities that may engage users.
Recent research has found that people assign high awareness and human resemblance to such agents \cite{scott2023you}. Furthermore, \citet{ross2023programmer} noted the high quality of generated replies and the agent's capacity to help users in certain domain activities (e.g., coding).

Qualitative outcomes by participants testing the LLM version of \systemName{} also pointed out that the chatbot was perceived as easy and interesting.
Additionally, given that the experimental task was executed by participants who were not experts in the specific environmental sustainability field, and generating good and reliable home automation represents a trivial task \cite{chen2022fix, huang2015supporting}, we can argue that LLM can enable more people (including non-experts) to set up routines to make their home consumption more optimal and sustainable.
Finally, a previous study by \citet{zhang2022promote} indicates that users' degree of involvement with a digital application is linked to their ecologically sustainable attitudes, which supports the results achieved by our application even if a proper effectiveness evaluation was not carried out.

However, it is worth noting that the study provided in this work was carried out in a laboratory setting with a hypothetical scenario, which has a substantial influence on the ecological validity of the experiences. Even though participants engaged with a functional conversational agent, the appliances' data was based on injected data from the available dataset, and users' actions did not affect real equipment. Furthermore, neither pre- or post-questionnaires were administered to participants about their environmental attitudes.

\section{Conclusion and Future Works}
\label{sec:conclusion}
We presented the potential of Large Language Models in advancing the capabilities of smart home automation systems to promote more sustainable household practices. Through a detailed comparative study and subsequent user evaluation of the \textit{EcoMate} application, it is evident that LLMs, particularly GPT models, have great abilities in generating effective home automation routines that can be parsed by the HomeAssistant framework.

The first evaluation revealed that GPT models, especially GPT4, excelled in producing valid JSON structures that were correctly executed by HomeAssistant. This ability showcases the effectiveness of these models in understanding complex user commands and transforming them into actionable automation routines. However, there remain challenges, notably in JSON formatting and message malformation, where the models often struggled with the correct syntax or produced malformatted outputs that could not be recognized by HomeAssistant.

Our second evaluation (a lab user study) further underscores the accessibility and likeability of the \textit{EcoMate} application, supported by GPT3.5. Participants positively rated the system on usability and likability, compared to a traditional rule-based one. Such results also reflect the system's potential to facilitate non-expert users in setting up and managing home automation for environmental sustainability.

The introduction of "green" prompts also provided subtle yet profound enhancements in the routines generated, promoting actions that align with sustainable energy practices. While the quantitative metrics indicated minimal differences between "green" and "no green" prompts, the qualitative analyses suggest a deeper integration of sustainability into the routines generated under green conditions.

These insights pave the way for further research into the optimization of LLMs for smart home environments. Such future research should focus on fine-tuning the models to improve their accuracy and assess their effectiveness in real-world scenarios (longitudinal and on-the-field studies). Additionally, exploring the integration of these technologies into more comprehensive smart home systems could further enhance their utility and adoption, providing a robust platform for promoting sustainable living through advanced technology.

\bibliographystyle{elsarticle-num-names} 
\bibliography{sample-base}

\appendix
\section{Comparative Study Prompt}
\label{app:comparison-promt}
This section contains the prompts used to compare the different LLMS.
Table \ref{tab:comparison-prompt-green} contains the "green" prompt, while Table \ref{tab:comparison-prompt-no-green} the "no green" one. The text \emph{emphasized} in italics wants to highlight the difference between the two prompts.

\begin{table}[H]
\centering
\caption{"green" Prompt}
\label{tab:comparison-prompt-green}
\begin{tabularx}{\linewidth}{lX}
\toprule 
\textbf{General Part} & You are \systemName{}. As a highly intelligent AI, you provide guidance on creating a HomeAssistant \emph{routine for sustainable energy practices, energy consumption optimization, and cultivating environmentally friendly habits}. You possess the above JSON containing information about your appliances and their energy consumption\emph{, and you aim to give advice and potentially generate HomeAssistant routines for energy management and energy saving.} \\
\midrule 
\textbf{Routine Part} & Your response should be concise and straight to the point. You are not allowed to talk about anything else. Provide a routine for Home Assistant and always provide JSON code for Home Assistant REST APIs. Always wrap the JSON of the routine inside \textasciigrave\textasciigrave\textasciigrave; do not generate YAML code. Generate a complete Home Assistant JSON code that sets up the routine using the JSON data I provide you about the appliances, the sensors of the house, and the consumption. \\
\midrule 
\textbf{Explanation Part} & After the answer, explain in a short way (maximum 20 words) the process you follow to generate the response. In particular, elaborate on the decision-making process for constructing a coherent and informative response.
Generate only one routine that "[USER COMMAND]" \\
\bottomrule
\end{tabularx}
\end{table}

\begin{table}[H]
\centering
\caption{"no green" Prompt}
\label{tab:comparison-prompt-no-green}
\begin{tabularx}{\linewidth}{lX}
\toprule 
\textbf{General Part} & You are \systemName{}. As a highly intelligent AI, you provide guidance on creating a HomeAssistant routine. You possess the above JSON, containing information about your appliances and their energy consumption.\\
\midrule 
\textbf{Routine Part} & Your response should be concise and straight to the point. You are not allowed to talk about anything else. Provide a routine for Home Assistant and always provide JSON code for Home Assistant REST APIs. Always wrap the JSON of the routine inside \textasciigrave\textasciigrave\textasciigrave; do not generate YAML code. Generate a complete Home Assistant JSON code that sets up the routine using the JSON data I provide you about the appliances, the sensors of the house, and the consumptions.\\
\midrule 
\textbf{Explanation Part} & After the answer, explain in a short way (maximum 20 words) the process you follow to generate the response. In particular, elaborate on the decision-making process for constructing a coherent and informative response. Generate only one routine that "[USER COMMAND]" \\
\bottomrule
\end{tabularx}
\end{table}

\section{Comparative Study Results}
\label{app:comparison-result}

\begin{table}[H]
\centering
\caption{ANOVA - JSON validity}
\label{table:anova_JSON}
\resizebox{\linewidth}{!}{
\begin{tabular}{lrrrrr}
\toprule
\multicolumn{1}{c}{\textbf{~}} & \multicolumn{1}{c}{\textbf{Sum of Squares}} & \multicolumn{1}{c}{\textbf{df}} & \multicolumn{1}{c}{\textbf{Mean Square}} & \multicolumn{1}{c}{\textbf{F}} & \multicolumn{1}{c}{\textbf{p}} \\
\midrule
LLM & 102.7996 & 5 & 20.5599 & 209.2066 &  $<$.001 \\
PROMPT & 0.0270 & 1 & 0.0270 & 0.2743 & 0.601 \\
TEMPERATURE & 9.95e-4 & 1 & 9.95e-4 & 0.0101 & 0.920 \\
LLM * PROMPT & 1.0862 & 5 & 0.2172 & 2.2106 & 0.051 \\
LLM * TEMPERATURE & 0.4781 & 5 & 0.0956 & 0.9730 & 0.433 \\
PROMPT * TEMPERATURE & 0.0815 & 1 & 0.0815 & 0.8296 & 0.363 \\
LLM * PROMPT * TEMPERATURE & 0.2151 & 5 & 0.0430 & 0.4378 & 0.822 \\
Residuals & 88.4481 & 900 & 0.0983 & ~ & ~ \\
\bottomrule
\end{tabular}
}
\end{table}

\begin{table}[H]
\centering
\caption{ANOVA - Latency}
\label{table:anova_latency}
\resizebox{\linewidth}{!}{
\begin{tabular}{lrrrrr}
\toprule
\multicolumn{1}{c}{\textbf{~}} & \multicolumn{1}{c}{\textbf{Sum of Squares}} & \multicolumn{1}{c}{\textbf{df}} & \multicolumn{1}{c}{\textbf{Mean Square}} & \multicolumn{1}{c}{\textbf{F}} & \multicolumn{1}{c}{\textbf{p}} \\
\midrule
LLM & 2.71e+10 & 5 & 5.41e+9 & 84.1159 & $<$.001 \\
PROMPT & 3.14e+6 & 1 & 3.14e+6 & 0.0488 & 0.825 \\
TEMPERATURE & 9.25e+7 & 1 & 9.25e+7 & 1.4375 & 0.231 \\
LLM * PROMPT & 1.96e+8 & 5 & 3.93e+7 & 0.6106 & 0.692 \\
LLM * TEMPERATURE & 3.28e+8 & 5 & 6.57e+7 & 1.0205 & 0.404 \\
PROMPT * TEMPERATURE & 5.89e+7 & 1 & 5.89e+7 & 0.9155 & 0.339 \\
LLM * PROMPT * TEMPERATURE & 1.33e+8 & 5 & 2.66e+7 & 0.4142 & 0.839 \\
Residuals & 5.79e+10 & 900 & 6.43e+7 & ~ & ~ \\
\bottomrule
\end{tabular}
}
\end{table}

\begin{table}[H]
\centering
\caption{ANOVA - Rel}
\label{table:anova_rel}
\resizebox{\linewidth}{!}{
\begin{tabular}{lrrrrr}
\toprule
\multicolumn{1}{c}{\textbf{~}} & \multicolumn{1}{c}{\textbf{Sum of Squares}} & \multicolumn{1}{c}{\textbf{df}} & \multicolumn{1}{c}{\textbf{Mean Square}} & \multicolumn{1}{c}{\textbf{F}} & \multicolumn{1}{c}{\textbf{p}} \\
\midrule
LLM & 6.43742 & 3 & 2.14581 & 4.06366 & 0.007 \\
PROMPT & 0.07583 & 1 & 0.07583 & 0.14360 & 0.705 \\
TEMPERATURE & 0.00475 & 1 & 0.00475 & 0.00899 & 0.924 \\
LLM * PROMPT & 0.16404 & 3 & 0.05468 & 0.10355 & 0.958 \\
LLM * TEMPERATURE & 0.56989 & 3 & 0.18996 & 0.35974 & 0.782 \\
PROMPT * TEMPERATURE & 0.12729 & 1 & 0.12729 & 0.24105 & 0.624 \\
LLM * PROMPT * TEMPERATURE & 2.05644 & 3 & 0.68548 & 1.29814 & 0.274 \\
Residuals & 291.48245 & 552 & 0.52805 & ~ & ~ \\
\bottomrule
\end{tabular}
}
\end{table}

\begin{table}[H]
\centering
\caption{ANOVA - Boolean difference}
\label{table:anova_bool_diff}
\resizebox{\linewidth}{!}{
\begin{tabular}{lrrrrr}
\toprule
\multicolumn{1}{c}{\textbf{~}} & \multicolumn{1}{c}{\textbf{Sum of Squares}} & \multicolumn{1}{c}{\textbf{df}} & \multicolumn{1}{c}{\textbf{Mean Square}} & \multicolumn{1}{c}{\textbf{F}} & \multicolumn{1}{c}{\textbf{p}} \\
\midrule
LLM & 2.163 & 5 & 0.4326 & 2.298 & 0.044 \\
TEMPERATURE & 0.158 & 1 & 0.1579 & 0.838 & 0.360 \\
LLM * TEMPERATURE & 0.437 & 5 & 0.0875 & 0.464 & 0.803 \\
Residuals & 84.169 & 447 & 0.1883 & ~ & ~ \\
\bottomrule
\end{tabular}
}
\end{table}

\begin{table}[H]
\centering
\caption{ANOVA - Latency difference}
\label{table:anova_latency_diff}
\resizebox{\linewidth}{!}{
\begin{tabular}{lrrrrr}
\toprule
\multicolumn{1}{c}{\textbf{~}} & \multicolumn{1}{c}{\textbf{Sum of Squares}} & \multicolumn{1}{c}{\textbf{df}} & \multicolumn{1}{c}{\textbf{Mean Square}} & \multicolumn{1}{c}{\textbf{F}} & \multicolumn{1}{c}{\textbf{p}} \\
\midrule
LLM & 1.57e+8 & 5 & 3.13e+7 & 0.494 & 0.781 \\
TEMPERATURE & 1.99e+8 & 1 & 1.99e+8 & 3.134 & 0.077 \\
LLM * TEMPERATURE & 5.14e+8 & 5 & 1.03e+8 & 1.622 & 0.153 \\
Residuals & 2.83e+10 & 447 & 6.34e+7 & ~ & ~ \\
\bottomrule
\end{tabular}
}
\end{table}

\begin{table}[H]
\centering
\caption{ANOVA - Similarity}
\label{table:anova_sim}
\resizebox{\linewidth}{!}{
\begin{tabular}{lrrrrr}
\toprule
\multicolumn{1}{c}{\textbf{~}} & \multicolumn{1}{c}{\textbf{Sum of Squares}} & \multicolumn{1}{c}{\textbf{df}} & \multicolumn{1}{c}{\textbf{Mean Square}} & \multicolumn{1}{c}{\textbf{F}} & \multicolumn{1}{c}{\textbf{p}} \\
\midrule
LLM & 0.2838 & 5 & 0.05677 & 9.910 & $<$.001 \\
TEMPERATURE & 0.0110 & 1 & 0.01104 & 1.927 & 0.166 \\
LLM * TEMPERATURE & 0.0257 & 5 & 0.00514 & 0.897 & 0.483 \\
Residuals & 2.5607 & 447 & 0.00573 & ~ & ~ \\
\bottomrule
\end{tabular}
}
\end{table}

\section{GPT3.5 Prompt}
\label{app:specific-design-promt}

\begin{table}[H]
\centering
\caption{Prompt used to feed the GPT3.5 model}
\label{tab:specific-design-promt}
\begin{tabularx}{\linewidth}{lX}
\toprule
\textbf{General Part} & You are \systemName{}. Address me as [USERNAME]. As a highly intelligent AI, you provide guidance on green energy practices, energy consumption optimization, and cultivating environmentally friendly habits. You possess JSON files containing information about your appliances and their energy consumption, and you aim to give advice and potentially generate HomeAssistant routines for energy management. Your response should be concise and straight to the point. You are not allowed to talk about anything else. If you're asked to provide a routine, do it for Home Assistant and always provide JSON code for Home Assistant RETS APIs; do not generate YAML code. \\
\midrule
\textbf{Routine Part} & Only if I ask you to explicitly create or generate a routine I want you to create a Home Assistant routine. Generate a complete Home Assistant JSON code that sets up the routine. You need to use the instructions I provided you to generate the routine. If not asking for a routine answer me in about my concerns. Just if needed, you can take into account the list of appliances I provide you. \\
\midrule
\textbf{Explanation Part} & After the answer, explain in a short way (maximum 20 words) the process you follow to generate the response. In particular, elaborate on the decision-making process for constructing a coherent and informative response. \\
\bottomrule
\end{tabularx}
\end{table}

\section{\systemName{} User Evaluation}
\label{app:user-evaluation}

\subsection{Scenario and Tasks}
\label{app:scenario-detail}

\subsubsection{Scenario}
\textit{Imagine living in a smart home with connected appliances. You are interested in reducing energy consumption and making the use of household appliances more sustainable. You recently discovered this application with a chatbot to create customized routines that automate the use of their appliances.}

\subsubsection{Tasks}
\begin{enumerate}
    \item Browse the app to see which appliances you have connected
    \item Interact with the chatbot to understand how much one of your appliances consumes
    \item Interact with the chatbot to discover useful green energy practices
    \item Create a new automation (i.e., routine) through the chatbot to improve the sustainability of your home
    \item Check the routine you just created    
\end{enumerate}

\subsubsection{Appliances}
Appliances were not presented to participants in the paper sheet but were directly visible in the specific tab of the \systemName{}.
Appliances are:
\begin{itemize}
    \item Smart TV
    \item Air Conditioner
    \item Oven
    \item Smart Lights
    \item Thermostat
\end{itemize}

\subsection{Questionnaire}
\label{app:questionnaire}

\begin{itemize}
    \item[] \underline{Demographics}
    \item Age
    \item Gender
    \item[] \underline{System Usability Scale (SUS) \cite{bangor2009determining}} 
    \item[] See the original paper \citet{brooke1996sus}
    
    \item[] \underline{Parasocial Interaction (PSI) scale \cite{tsai2021chatbots}} 
    \item[] \textit{Perceived Dialogue}
    \item I felt like I was engaged in an active dialogue with the chatbot
    \item My interactions with the chatbot felt like a back and forth conversation
    \item The chatbot responded quickly to my inputs and requests
    \item The chatbot was efficient in responding to my activities
    \item[] \textit{User Engagement}
    \item Time appeared to go by very quickly when I was interacting with the website
    \item I spent more time on the website than I had intended
    \item While I was interacting with the website, I was able to block out most other distractions
    \item While I was interacting with the website, I was immersed in what I was doing
    \item I lost track of time when I was interacting with the website
    \item[] \textit{Interaction Satisfaction}
    \item I enjoyed my interaction with the website
    \item Interacting with the website is satisfying
    \item Interacting with the website is frustrating $\dag$
    \item I think the website gave me a headache $\dag$
    \item Interacting with the website is very awkward $\dag$
\end{itemize}

Please note that items marked with \emph{$\dag$} are the reverse-coded ones.

\end{document}